\documentclass{aa}

\usepackage{natbib}
\bibpunct{(}{)}{;}{a}{}{,} 

\usepackage{lscape}

\usepackage{placeins}
\usepackage{graphicx}
\usepackage{amsmath}
\usepackage{txfonts}

\graphicspath{{Figures/}}

\newcommand{\mdisk}[0]{\ensuremath{M_{\rm disk}}}
\newcommand{\mgas}[0]{\ensuremath{M_{\rm gas}}\ }
\newcommand{\mdust}[0]{\ensuremath{M_{\rm dust}}\ }
\newcommand{\CO}[0]{\ensuremath{^{12}\mathrm{CO}}\ }
\newcommand{\gdratio}[0]{\ensuremath{\Delta_{\rm gd}}}
\newcommand{\dCO}[0]{\ensuremath{\delta_{\rm C,O}}\ }
\newcommand{\ratgasdust}[0]{\ensuremath{R_{\rm 90, gas}/R_{\rm 90, dust}}}
\newcommand{\rgas}[0]{\ensuremath{R_{\rm 90, gas}}}
\newcommand{\rdust}[0]{\ensuremath{R_{\rm 90, dust}}}
\newcommand{\rc}[0]{\ensuremath{R_{\rm c}}}
\newcommand{\rco}[0]{\ensuremath{R_{\rm CO\ disk}}}

\newcommand{\alp}[0]{\ensuremath{\alpha_{\rm turb}}}

\usepackage{color}
\usepackage{subcaption}

\begin{document} 

   \title{Gas vs dust sizes of protoplanetary disks: effects of dust evolution}

   \author{L. Trapman \inst{1}
          \and
          S. Facchini \inst{2,3}
          \and
          M.R. Hogerheijde \inst{1,4}
          \and
          E.F. van Dishoeck \inst{1,2}
          \and 
          S. Bruderer \inst{2}
          }

   \institute{
            Leiden Observatory, Leiden University, Niels Bohrweg 2, NL-2333 CA Leiden, The Netherlands \\
            \email{trapman@strw.leidenuniv.nl}
        \and
            Max-Planck-Institute f\"{u}r Extraterrestrische Physik, Giessenbachstra{\ss}e, D-85748 Garching, Germany
        \and
            European Southern Observatory, Karl-Schwarzschild-Str. 2, D-85748 Garching bei M\"unchen, German
        \and
            Anton Pannekoek Institute for Astronomy, University of Amsterdam, Science Park 904, 1090 GE Amsterdam, The Netherlands
             }
   \date{Received 17 December 2018; Accepted 13 March 2019}

  \abstract
   {The extent of the gas in protoplanetary disks is observed to be universally larger than the extent of the dust. This is often attributed to radial drift and grain growth of the mm grains, but line optical depth produces a similar observational signature.}
   {We investigate in what parts of the disk structure parameter space dust evolution and line optical depth are the dominant drivers of the observed gas and dust size difference.}
   {Using the thermochemical model \texttt{DALI} with dust evolution included we ran a grid of models aimed at reproducing the observed gas and dust size dichotomy.}
  {The relation between $R_{\rm dust}$ and dust evolution is non-monotonic and depends on the disk structure. $R_{\rm gas}$ is directly related to the radius where the CO column density drops below $10^{15}\ \mathrm{cm}^{-2}$ and CO becomes photodissociated.
  $R_{\rm gas}$ is not affected by dust evolution but scales with the total CO content of the disk. $R_{\rm gas}/R_{\rm dust} > 4$ is a clear sign for dust evolution and radial drift in disks, but these cases are rare in current observations. For disks with a smaller $R_{\rm gas}/R_{\rm dust}$, identifying dust evolution from $R_{\rm gas}/R_{\rm dust}$ requires modelling the disk structure including the total CO content. To minimize the uncertainties due to observational factors requires FWHM$_{\rm beam} < 1\times$ the characteristic radius and a peak SNR $ > 10$ on the $^{12}$CO emission moment zero map. For the dust outer radius to enclose most of the disk mass, it should be defined using a high fraction (90-95\%) of the total flux. For the gas, any radius enclosing $> 60\%$ of the $^{12}$CO flux will contain most of the disk mass.}
   {To distinguish radial drift and grain growth from line optical depth effects based on size ratios requires disks to be observed at high enough angular resolution and the disk structure should to be modelled to account for the total CO content of the disk.}
   
    \keywords{Protoplanetary disks -- Astrochemistry -- Molecular processes -- Radiative transfer -- Line: formation -- Methods: numerical}   

   \maketitle
%

\section{Introduction}
\label{sec: introduction}

Observations of exoplanetary systems have revealed that they come in a large range of sizes, from multiple planets in the central $\sim0.3-1$ AU of the system (e.g. TRAPPIST-1 or Kepler-90, \citealt{Gillon2017,Cabrera2014}) to systems with Jupiter mass planets in $\sim70$ AU orbits around their host star (HR 8799, \citealt{marois2008}). The diversity in planetary systems is linked to the diversity in protoplanetary disks from which these planets have formed.

To better understand this link, measurements of the properties of these disks are required. Of particular interest are the spatial extent of the gas, which sets the evolution of the disk, and the behaviour of the millimetre (mm) grains, that form the building blocks of the planets. 

Observations have shown that the gas, traced by the $^{12}$CO emission, extends further out than the mm grains traced by the (sub)mm continuum emission. Two physical processes contribute to the observed size dichotomy. The first is a difference in optical depth, with the line optical depth being much higher than the continuum optical depth (e.g., \citealt{Dutrey1998,GuilloteauDutrey1998,Facchini2017}). Depending on how rapidly the density profile drops off in the outer disk, the optically thin continuum emission will drop below the detection limit before the optically thick $^{12}$CO emission. Based on the self-similar solution of viscous evolution, an exponentially tapered power law profile has been proposed to simultaneously fit the extent of the gas emission and the extent of the dust emission (e.g., \citealt{Hughes2008,Andrews2009}, see also \citealt{Panic2008}). 

The second physical process setting the observed size dichotomy is grain growth and the subsequent inward radial drift of mm-sized grains. There is already extensive observational evidence that grains can grow to at least mm sizes (e.g., \citealt{Testi2003, Natta2004,Lommen2007,AndrewsWilliams2005,AndrewsWilliams2007a,Ricci2010}).
The large grains have decoupled from the gas. The gas is partly supported by a pressure gradient and therefore moves at slightly sub-Keplerian velocities. Gas drag causes the large grains, moving at Keplerian velocities, to slow down and move inward. In addition, the maximum grain size seems to be decreasing as function of radial distance from the star, supported by both observations (e.g., \citealt{Guilloteau2011,Perez2012,Perez2015,Menu2014,Tazzari2016,Tripathi2018}) and theoretical modelling results (e.g., \citealt{Birnstiel2010,Birnstiel2012}).

Both radial drift and radially dependent grain growth cause the mm-sized grains to be confined in the inner regions of the disk, resulting in compact continuum emission at millimetre wavelengths. 
Dust evolution also affects the CO chemistry and gas temperature and could therefore also change the observed gas disk size, determined from the $^{12}$CO emission. \cite{Facchini2017} found that grain growth and settling results in colder gas with respect to the dust at intermediate disk heights, which reduces the CO excitation and emission. 

The Atacama Large Millimeter/submillimeter Array (ALMA) is transforming our understanding of disk sizes. High resolution observations have shown that for several disks the dust outer edge drops off too sharply with radius and cannot be explained with the same exponential taper that reproduces the $^{12}$CO emission (e.g. \citealt{Andrews2012,Andrews2016,deGregorioMonsalvo2013,Pietu2014,Cleeves2016}). 

In addition, complete surveys of disks with ALMA have made it possible to study these disk properties not only for individual disks but also for the full disk population (e.g. Taurus: \citealt{Andrews2013,WardDuong2018}, Lupus: \citealt{ansdell2016,ansdell2018}, Chamaeleon I: \citealt{Pascucci2016,Long2017}, Upper Sco: \citealt{Barenfeld2016,Barenfeld2017}, $\sigma$ Ori: \citealt{ansdell2017}, IC 348: \citealt{Ruiz-Rodriguez2018} ). One of the main findings of these surveys is that most disks have very compact mm emission. For example, in the Lupus Survey \citep{ansdell2016,ansdell2018} $\sim45\%$ remain unresolved at $\sim 20$ AU radius resolution. It should be noted that the integration time used by these surveys is short, 1-2 min per source, resulting in a low signal-to-noise ratio (SNR) on the gas lines. 

Observations show that the outer radius of the gas disk, traced by the \CO emission, is universally larger than the mm dust disk, as traced by the mm emission. \cite{ansdell2018} measured gas and dust outer radii for 22 disks from the \CO 2-1 emission and 1.3 mm continuum emission. They found gas-dust size ratios $R_{\rm gas}/R_{\rm dust}$ ranging from 1.5-3.5, with an average $\langle R_{\rm gas}/R_{\rm dust}\rangle = 1.96 \pm 0.04|_{\sigma_{\rm obs}}$. Larger gas-dust ratios ($R_{\rm gas}/R_{\rm dust} > 4$) have been found for a few individual disks (e.g., Facchini et al., subm.).
It should be noted measurements of disk sizes are biased toward the most massive disks. Gas-dust sized differences for the faint end of the disk population are not well explored with current sensitivities and angular resolutions. 

Both optical depth and radial drift contribute to the observed gas-dust size ratio. Quantitative comparison of how much these two effects affect the gas-dust size ratio has so far been limited to a single disk structure  appropriate for the large and massive disk HD~163296 \citep{Facchini2017}.

In this paper, we expand the quantitative analysis of optical depth to a range of disk structures including dust growth and radial drift. In particular we focus on how disk mass, disk size and dust evolution affect the gas-dust size difference. Additionally, we address what role observational factors like resolution and sensitivity play in the observed gas-dust size difference. The setup of the method and the models used in the paper are described in Section \ref{sec: models}. The results are presented in Section \ref{sec: results}. In Section \ref{sec: discussion} the connection between dust evolution and the dust outer radius is discussed. The conclusions are presented in Section \ref{sec: conclusions}.


\section{Models}
\label{sec: models}

\subsection{\texttt{DALI} with dust evolution}
\label{sec: DALI with dust evolution}

To study the effects of radial drift, grain growth and optical depth on the gas-dust size dichotomy we use the thermo-chemical model \texttt{DALI} \citep{Bruderer2012,Bruderer2013} with dust evolution included by \cite{Facchini2017}.

For a given physical structure, this version of \texttt{DALI} first calculates the radial dependence of the grain size distribution following the reconstruction routine from \cite{Birnstiel2015}. This semi-analytical prescription provides a good representation of the more complete numerical models in \cite{Birnstiel2010}. They divide the dust in the disk into two regimes: In the inner part of the disk, dust evolution is fragmentation dominated and the maximum grain size is set by the fragmentation barrier. In the outer disk, the maximum grain size is set by radial drift. The dust evolution is run for 1 Myr. Tests with models run for 10 Myr showed that the dust evolution timescale has only minimal effect on the dust outer radius (less than 17 \%).

Next, dust settling is calculated by solving the advection-diffusion equation in the vertical direction for each grain size bin at every radial point in the model. Opacities are calculated at each $(r,z)$ point of the model using the resulting local grain size distribution. 

It should be noted here that the local gas-to-dust mass ratio ($\Delta_{\rm gd}$) in the models is kept fixed at $\gdratio = 100$, i.e., only the dust properties are changed.

The \texttt{DALI} thermo-chemical computation can be split into three consecutive steps: First the continuum radiative transfer equation is solved using the input stellar spectrum and the grain opacities calculated in the previous step. This is done using a 3D Monte Carlo method. Next the abundances of atomic and molecular species are calculated by solving the time dependent chemistry at each point in the model. In this step the local grain size distribution is taken into account when computing the dust surface area available for processes such as gas-grain collisions, H$_2$ formation rate, freeze-out, thermal and non-thermal desorption, and hydrogenation. Using a non-LTE formulation the excitation of levels of the atomic and molecular species are calculated and the resulting gas temperature is determined by balancing the heating and cooling processes. Both the chemistry and the excitation are temperature dependent. The calculation is therefore performed iteratively until a self-consistent solution is found. A more detailed description of \texttt{DALI} can be found in Appendix A of \cite{Bruderer2013}. The implementation of dust evolution in \texttt{DALI} is described in \cite{Facchini2017}.

\subsection{Model setup}
\label{sec: model setup}

The gas surface density profile of the models is described by a tapered power law that is often used to describe protoplanetary disks (e.g., \citealt{Hughes2008,Andrews2009,Andrews2011,Tazzari2017}). This simple parametric structure is based on the assumption that the gas structure is set by viscous accretion, where $\nu \propto R^{\gamma}$ \citep{LyndenBellPringle1974,Hartmann1998}

\begin{equation}
\label{eq: tapered powerlaw}
\Sigma_{\rm gas} (R)  = \frac{\mdisk (2-\gamma)}{2\pi \rc^2} \left( \frac{R}{\rc} \right)^{-\gamma} \exp \left[-\left(\frac{R}{\rc}\right)^{2-\gamma}\right].
\end{equation}

\noindent Here \rc\ is the characteristic radius where the surface density profile transitions from a power law to an exponential taper. 

Under the assumption of vertical isothermality and hydrostatic equilibrium the vertical structure is given by a Gaussian density distribution \citep{KenyonHartmann1987}

\begin{equation}
\label{eq: density structure}
\rho_{\rm gas} = \frac{\Sigma_{\rm gas}}{\sqrt{2\pi} R h} \exp \left[ - \frac{1}{2} \left( \frac{z}{Rh} \right)^2 \right],
\end{equation}
where $h = h_c (R/\rc)^{\psi}$, $\psi$ is the flaring powerlaw index and $h_c$ is the disk opening angle at \rc.

\subsection{Grid of models}
\label{sec: grid of models}

Both the characteristic size \rc\ and the total disk mass $M_{\rm disk}$ are expected to affect the observed extent of the disk. A set of models was run varying both parameters: $\rc = 20, 50$ AU and $M_{\rm disk} = 10^{-2}, 10^{-3}, 10^{-4}, 10^{-5}\ \mathrm{M}_{\odot}$. No models with larger \rc\ were run as we aim to reproduce the bulk of disk population, most of which are found to be small. For each of these physical structures three models are run with $\alpha = 10^{-2}, 10^{-3}, 10^{-4}$. For reference, a model with the same (\rc, \mdisk) is run using DALI without dust evolution (\textit{no drift}). In this model the dust is split into two grain populations: small grains with sizes ranging between $50\ \AA$ and $1\ \mu$m and large grains with sizes between $1\ \mu$m and 1 mm. 
These large grains are restricted to a scale height of $\chi h$, with $\chi < 1$, simulating that these grains have settled towards the midplane. The mass ratio between the large and the small grains is given by $f_{\rm large}$. 

Standard volatile [C]/[H] $= 1.35\cdot10^{-4}$ and [O]/[H] $= 2.88\cdot10^{-4}$ are assumed in all models and the chemistry is evolved over a timescale of 1 Myr, which is a representative age for protoplanetary disks. For longer timescales, CO is converted into CH$_4$/C$_2$H$_2$, as shown in \cite{Bosman2018b} (see also \citealt{Schwarz2018,Dodson-Robinson2018} ). This results in a overall underabundance of volatile CO, which has been found in a number of disks (e.g., \citealt{Favre2013,Kama2016,Cleeves2016,McClure2016,miotello2017}). To investigate how such an underabundance in CO affects the observed gas disk size, a subset of models was run with a lower [C]/[H] and [O]/[H]. These models are discussed in Section \ref{sec: CO depletion}.

T Tauri stars are expected to have excess UV radiation as a result of accretion onto the stellar surface. This UV radiation is added to the spectra as a blackbody with T = 10000 K, with a luminosity computed from the accretion rate assuming that the gravitational potential energy is released as radiation with 100\% efficiency (see also \citealt{Kama2016a}).

For analysis, the disks are assumed to be face on ($i = PA = 0^{\circ}$). The effect of inclination is discussed in Appendix \ref{app: inclination} and is found to be minimal for $i\leq 50^{\circ}$. In total 32 models are run. Their parameters are found in Table \ref{tab: model parameters}.

\begin{table}[htb]
  \centering   
  \caption{\label{tab: model parameters}\texttt{DALI} parameters of the physical model.}
  \begin{tabular*}{0.8\columnwidth}{ll}
    \hline\hline
    Parameter & Range\\
    \hline
     \textit{Chemistry}&\\
     Chemical age & 1 Myr\\
     {[C]/[H]} & $1.35\cdot10^{-4}$\\
     {[O]/[H]} & $2.88\cdot10^{-4}$\\
     \textit{Physical structure} &\\ 
     $\gamma$ &  1.0\\ 
     $\psi$ & 0.1\\ 
     $h_c$ &  0.1 rad\\ 
     \rc & [20, 50] AU\\
     $M_{\mathrm{gas}}$ & [$10^{-5}$,$10^{-4}$, \\
                        &  $10^{-3}$,$10^{-2}$] M$_{\odot}$ \\
     Gas-to-dust ratio & 100 \\
     \textit{Dust properties} - \textit{no drift} &\\
     $f_{\mathrm{large}}$ & 0.85 \\
     $\chi$ & 0.2 \\
     \textit{Dust properties} - dust evolution   &\\
     \alp & [$10^{-2}$,$10^{-3}$,$10^{-4}$] \\
     $\rho_{\rm gr}$ & 2.5 g cm$^{-3}$ \\
     v$_{\rm frag}$ & 10 m s$^{-1}$\\
     composition & standard ISM$^{1}$\\
     \textit{Stellar spectrum} & \\
     $T_{\rm eff}$ & 4000 K + Accretion UV \\
     $L_{*}$ & 0.5 L$_{\odot}$  \\
     \textit{Observational geometry}&\\
     $i$ & 0$^{\circ}$ \\
     PA & 0$^{\circ}$ \\
     $d$ & 150 pc\\
    \hline
  \end{tabular*}
  \captionsetup{width=.75\columnwidth}
  \caption*{\footnotesize{$^{1}$\citealt{WeingartnerDraine2001}, see also Section 2.5 in \citealt{Facchini2017}. }}
\end{table}

\subsection{Measuring the outer radius}
\label{sec: measuring the outer radius}

To investigate the gas-dust size difference we have to measure the size of a disk from observations. A disk size metric that is often adopted for these purposes makes use of the cumulative intensity profile, i.e., the flux is measured in increasingly larger apertures. The outer radius ($R_{\rm 90}$) is defined as the radius that encloses 90\% of the total flux ($F_{\rm tot}$) of the disk
\begin{equation}
\label{eq: outer radius}
 0.9 = \frac{2\pi}{F_{\rm tot}} \int^{R_{\rm 90}}_{0} I_{\nu}(r')r' \mathrm{d}r'.
\end{equation}

This method has the advantage that it can be easily and homogeneously applied to a large number of disks, even if these disks show signs of substructure (see, e.g., \citealt{Tripathi2017, ansdell2018, andrews2018}). In addition, the method can be applied to the short integration observations used in recent surveys, where the limited sensitivity hinders a more complex analysis.

It should be noted that the resulting outer radius is an observational outer radius. How well this observational radius is related to underlying physical size of the disk is examined in Section \ref{sec: mass fraction vs flux fraction}. 

In this work, the gas outer radius is measured from the extent of the $^{12}$CO 2-1 emission in the moment zero map and the dust outer radius is measured from the extent of the 1300 $\mu$m continuum emission. Gas outer radii measured using the $^{12}$CO 3-2 emission differ from those measured using the $^{12}$CO 2-1 by less than 10\%. For comparison, gas outer radii measured instead from the $^{13}$CO 2-1 emission are shown in Appendix \ref{app: 13CO outer radii}.

Note that the moment zero map is a velocity integrated intensity (in Jy/beam km/s). This puts additional weight at the centre of the disk, where the line widths are larger. More comparable to the continuum emission would be to use the peak intensity map, defined as the peak intensity of the spectrum at each spatial point. In Appendix \ref{app: moment 0 vs moment 8} we compare gas outer radii derived from the moment 0 and the peak intensity map and find them to be nearly identical.

\section{Results}
\label{sec: results}

In this section we investigate how dust evolution shapes the continuum and line emission and how it affects the dust and gas outer radii. This effect is quantitatively compared to the influence of other disk parameters (\mdisk, \rc) and observational factors (signal-to-noise, size of the beam). 

\subsection{Dust radial intensity profiles}
\label{sec: dust radial profiles}

\begin{figure}
    \centering
    \includegraphics[width=\columnwidth]{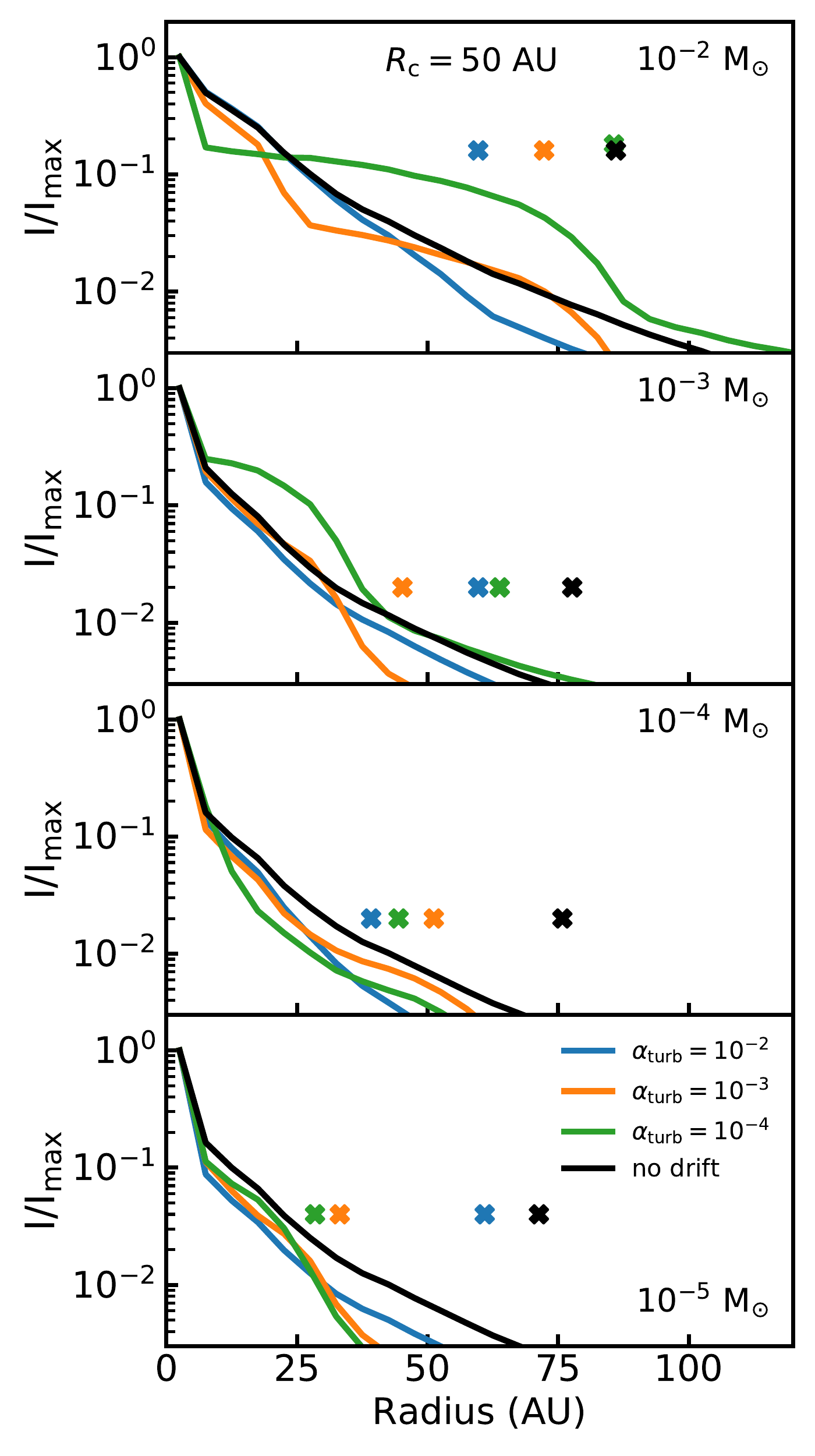}
    \caption{\label{fig: dust profiles} Radial intensity profiles of the 1.3 mm continuum emission, normalised to the peak intensity, for the models with \rc\ = 50 AU. Crosses at an arbitrary height above the line denote the dust outer radii, defined here as the radii enclosing 90\% of the total flux. The resulting cumulative intensity curves are shown in Figure \ref{fig: raw dust curves}.}
\end{figure}

Dust evolution changes the distribution of the grains responsible for the mm continuum emission, resulting in a different radial intensity profile. Through Equation \eqref{eq: outer radius} the dust outer radius is affected by the radial profile. In this section we investigate this link. 

Figure \ref{fig: dust profiles} shows the normalised dust radial profiles of the models with \rc\ = 50 AU on a logarithmic intensity scale. The radial profiles of the models with \rc\ = 20 AU are shown in Figure \ref{fig: raw dust profiles rc 20} in Appendix \ref{app: dust profiles rc 20}. Starting with the lowest disk mass ($\mdisk = 10^{-5}\ \mathrm{M}_{\odot}$), all intensity profiles fall off steeply in the inner 10 AU. For disks with $\alp = 10^{-3}-10^{-4}$ the intensity profile flattens between 10 and 25 AU and then steepens again. The dust outer radii of these two models are located at the second steepening off. For the model with $\alp = 10^{-2}$ and the \textit{no drift} model, the intensity profile is more extended and the dust outer radii of these models are larger. 

For the disks with $\mdisk = 10^{-4}\ \mathrm{M}_{\odot}$ all intensity profiles of the dust evolution models are similar within 25 AU and their outer radii therefore lie close together. 

At higher disk mass ($\mdisk \geq 10^{-3}\ \mathrm{M}_{\odot}$) and low viscosity ($\alp \leq 10^{-3}$) a plateau of emission can be seen. The prominence of this plateau increases as \alp\ decreases. For $\mdisk = 10^{-2}\ \mathrm{M}_{\odot}$ and $\alp = 10^{-4}$ about 75\% of the emission is in the plateau and it has a large effect on the cumulative flux and the location of \rdust\ (cf. Figure \ref{fig: raw dust curves}). 

The emission plateau is directly linked to the presence of mm grains in the outer disk. When \alp\ is low, the timescale for collisions that result in fragmentation is longer than the drift timescale and the size of the grains in the outer disk is set by radial drift (cf. \citealt{Birnstiel2015}). This causes a pile up of mm grains in the outer parts of the disk. Thus, the shape of the intensity profile is affected in a complex way by dust evolution. 

\subsection{$^{12}$CO radial intensity profiles}
\label{sec: gas radial profiles}

\begin{figure}
    \centering
    \includegraphics[width=\columnwidth]{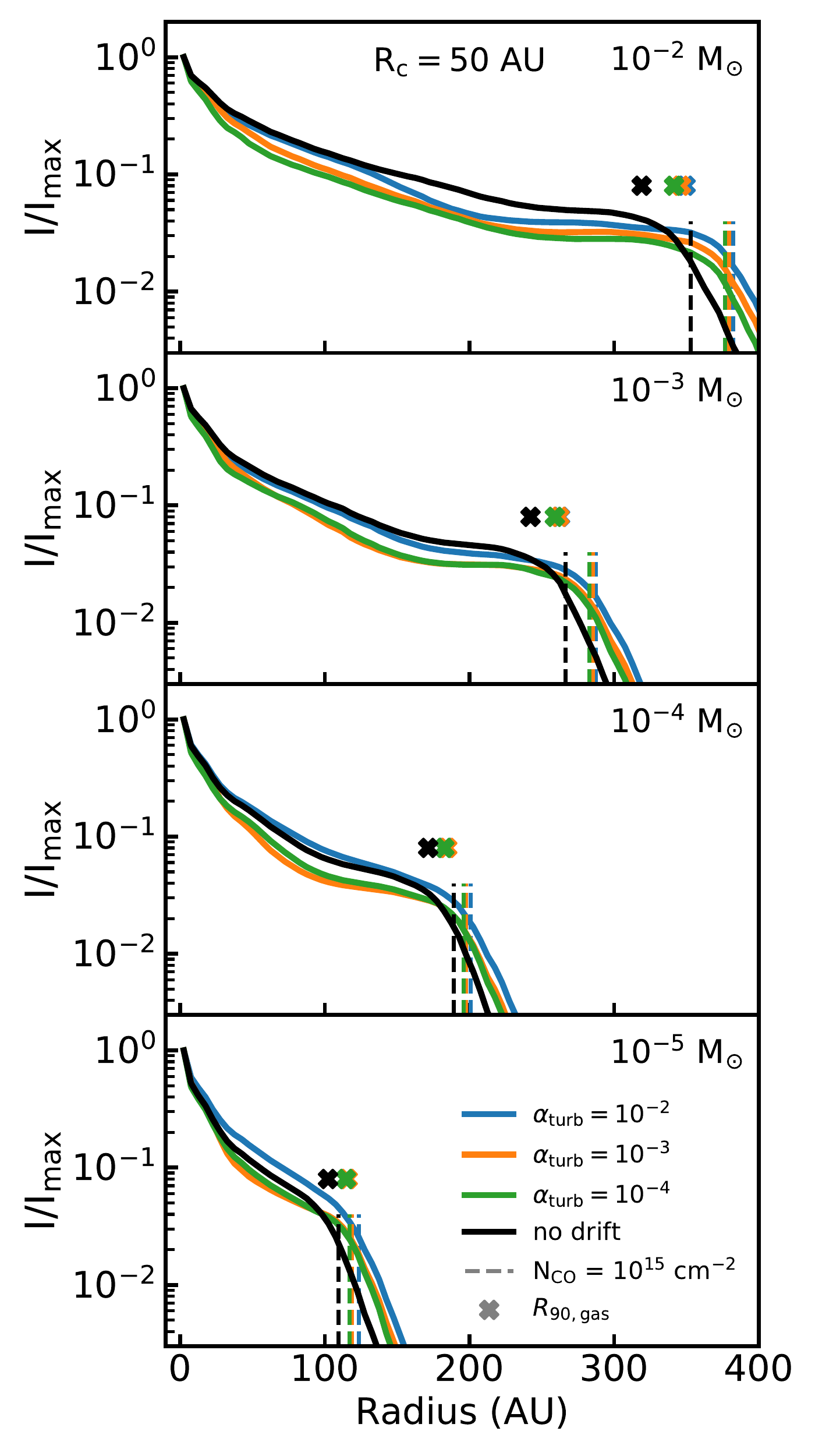}
    \caption{\label{fig: gas profiles} Radial intensity profiles of the $^{12}$CO 2-1 line emission, normalised to the peak intensity, for the models with \rc\ = 50 AU. Crosses above the line denote the dust outer radii, defined here as the radii enclosing 90\% of the total flux. Vertical dashed lines denote the radius at which the CO column density drops below $10^{15}\ \mathrm{cm}^{-2}$, where it can be photodissociated effectively.  }
\end{figure}

The gas outer radius is measured from $^{12}$CO 2-1 line emission, which is expected to be mostly optically thick throughout the disk. Dust evolution could affect the $^{12}$CO emission by altering the temperature structure \citep{Facchini2017}. 

In Figure \ref{fig: gas profiles} the $^{12}$CO 2-1 line emission profiles of the models with \rc\ = 50 AU are examined. Within each mass bin, the profiles are very similar in shape, suggesting that the effect of dust evolution on the $^{12}$CO emission is neglible. The emission profiles drop off relatively slowly, which is expected for optically thick line emission that follows the temperature profile. At a certain radius the emission profile drops off steeply. In the model this radius corresponds to where the CO column ($N_{\rm CO}$) density drops below $10^{15}\ \mathrm{cm}^{-2}$. Below a CO column density of $N_{\rm CO} \leq 10^{15}\ \mathrm{cm}^{-2}$, CO is no longer able to effectively self-shield against photodissociation and is quickly removed from the gas phase \citep{vanDishoeckBlack1988}. Defining the radius at which $N_{\rm CO} = 10^{15}\ \mathrm{cm}^{-2}$ as \rco, this radius effectively encloses all of the CO emission as well as all of the volatile CO in the disk.

Using simple arguments, the observed gas outer radius \rgas\ can be related analytically to \rco. Assuming that the CO emission is optically thick ($I_{\rm CO} \sim T_{\rm gas}(R)\propto R^{-\beta}$) and that \rco encloses all $^{12}$CO flux, we can write (full derivation can be found in Appendix \ref{app: rgas-rco derivation}) 
\begin{equation}
\label{eq: rgas and rco relation}
\rgas = 0.9^{\frac{1}{2-\beta}}\rco = f^{\frac{1}{2-\beta}}\rco,
\end{equation}
where $f$ represents a more general case where the gas outer radius is defined using a flux fraction $f$.

\begin{figure}
    \centering
    \includegraphics[width=\columnwidth]{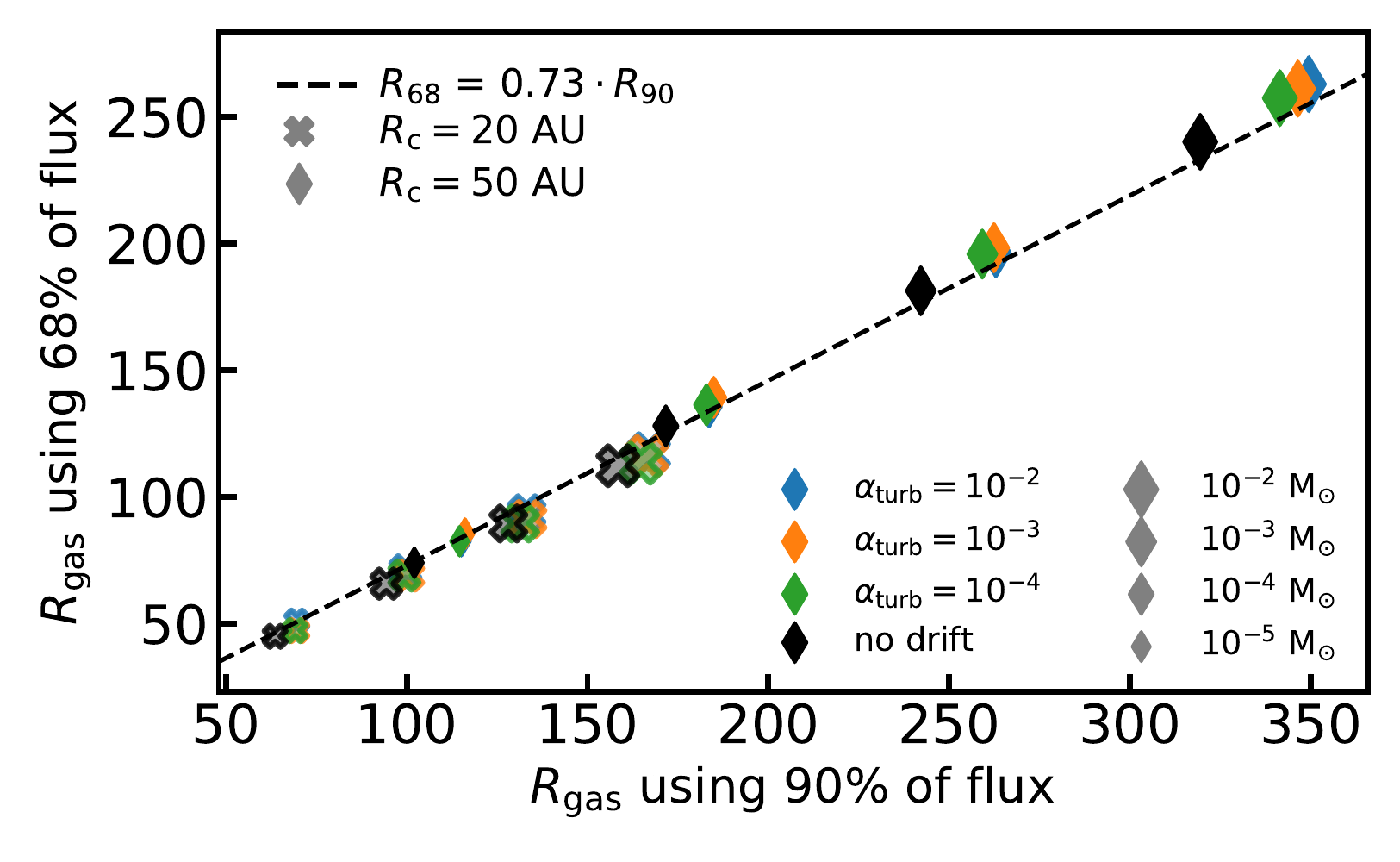}
    \caption{\label{fig: R90 vs R68 gas} Comparison between gas outer radii calculated using 90\% of the flux and using 68\% of the flux. Dashed line shows the expected relation between these two observational outer radii based on Equation \eqref{eq: rgas and rco relation}.} 
\end{figure}

Based on Equation \eqref{eq: rgas and rco relation} the fraction of flux $f$ used to define the gas outer radius should not affect the dependence of $R_{\rm f, gas}$ on disk parameters such as \mgas\ or \rc. To highlight this point, Figure \ref{fig: R90 vs R68 gas} compares gas outer radii defined using 90\% and 68\% of the total flux. Independent of \mgas, \rc\ and \alp\ the models follow a tight linear relation that matches the expected relation $R_{\rm 68, gas} = 0.73\rgas$, based on Equation \eqref{eq: rgas and rco relation}. 

Summarising, the observational gas outer radius is directly related to the point the the disk where the CO column density reaches $N_{\rm CO} = 10^{15}\ \mathrm{cm}^{-2}$. This relation is independent of the flux fraction used to define the gas outer radius.

\subsection{Effect of dust evolution on \rdust\ and \rgas}
\label{sec: alpha vs rgas and rdust}

In the previous section dust evolution was shown to change the continuum intensity profile and the dust outer radius. Dust evolution also affects the CO chemistry \citep{Facchini2017} and could therefore change the gas outer radius.

\begin{figure}
    \centering
    \includegraphics[width=\columnwidth]{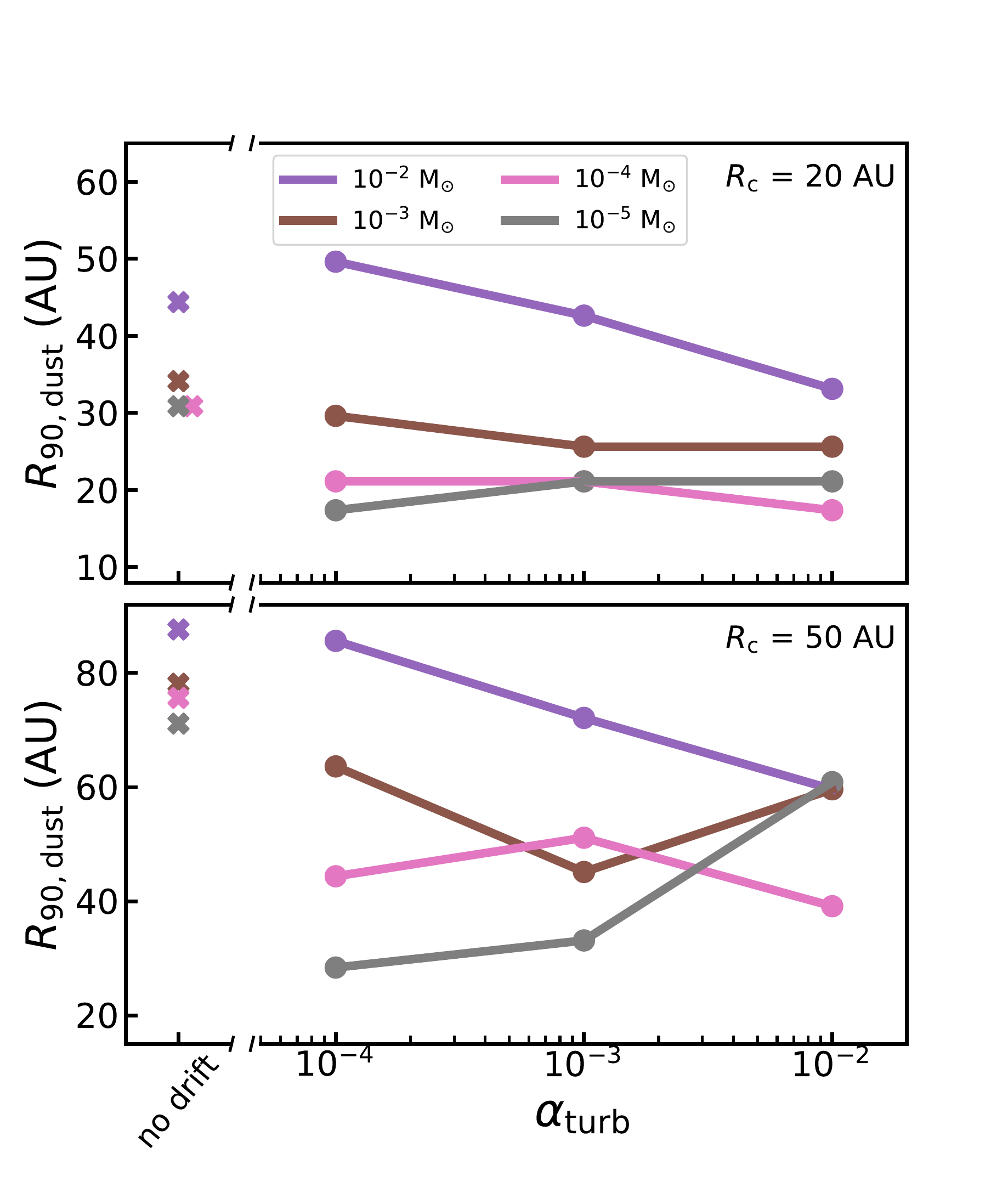}
    \caption{\label{fig: dust outer radii vs alpha}$R_{\rm 90, dust}$ as function \alp\ for models with different (\mdisk, \rc). Top panel shows models with $\rc = 20$ AU. Bottom panel shows models with $\rc = 50$ AU. The crosses show the $R_{\rm 90, dust}$ for \textit{no drift} model.}
\end{figure}

Figure \ref{fig: dust outer radii vs alpha} shows gas and dust outer radii as function of the turbulent $\alpha$ for different points in our (\mdisk-\rc) parameter space. Compared to \textit{no drift} model, the dust radii of the dust evolution models are smaller up to a factor 1.5. No obvious trend of $R_{\rm dust}$ with \alp\ is found. The dust outer radius scales with $\alpha_{\rm visc}$ at high disk mass of $10^{-2}$ M$_{\odot}$, corresponding to a disk dust mass of $10^{-4}$ M$_{\odot}$. The trend is negative, with a higher $\alpha$ corresponding to a smaller \rdust. At high $\alpha$ fragmentation sets the maximum grain size throughout the disk, preventing millimetre grains from forming.

For lower disk masses the behaviour of \rdust\ as function of \alp\ depends on the characteristic size of the disk \rc. The intensity profile of disks with $\rc = 20$ AU is dominated by an inner core that is largely unaffected by \alp. As a result \rdust\ remains approximately constant with \alp. For the larger disks ($\rc = 50$ AU) \rdust\ varies with \alp, but the trends are not monotonic for $\mdisk = 10^{-3} - 10^{-4}$ M$_{\odot}$. For $\mdisk = 10^{-5}$ the trend is monotonic again, but now \rdust\ is larger for higher \alp. 

\begin{figure}
    \centering
    \includegraphics[width=\columnwidth]{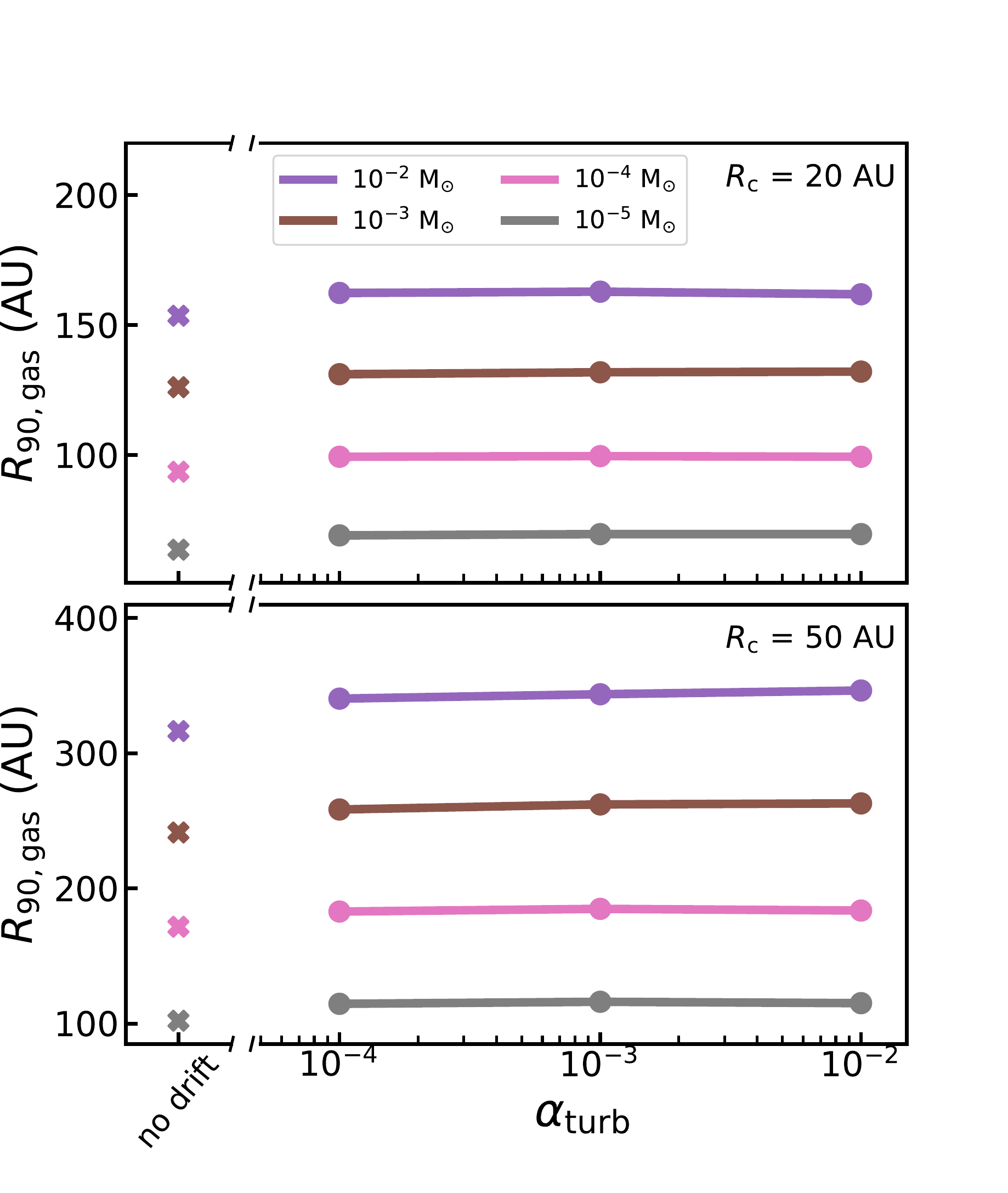}
    \caption{\label{fig: gas outer radii vs alpha}$R_{\rm 90, gas}$ as function $\alpha$ for models with different (\mdisk, \rc). Top panel shows models with $\rc = 20$ AU. Bottom panel shows models with $\rc = 50$ AU. The crosses show the $R_{\rm 90, gas}$ for the \textit{no drift} model. }
\end{figure}

The effect of dust evolution on the gas outer radius is shown in Figure \ref{fig: gas outer radii vs alpha}. There are no noticeable changes in \rgas\ when \alp\ is varied, indicating that \rgas\ is unaffected by dust evolution. The gas radii of the dust evolution models are larger than the gas radius of the \textit{no drift} model. 

A possible explanation is a difference in the amount of small grains in the outer disk. In the \textit{no drift} model a fixed fraction of the dust is in small grains. In the dust evolution model the maximum grain size decreases with radius, representing the larger grains drifting inward. However in our model framework no mass is actually transferred inward. As a result the amount of small grains in the outer disk is enhanced. These small grains can help shield the CO against photodissociation, allowing it to exist further out in the disk.  

\subsection{The effect of disk mass on \rgas\ and \rdust}
\label{sec: the effect of disk mass}

\begin{figure}
    \centering
    \includegraphics[width=\columnwidth]{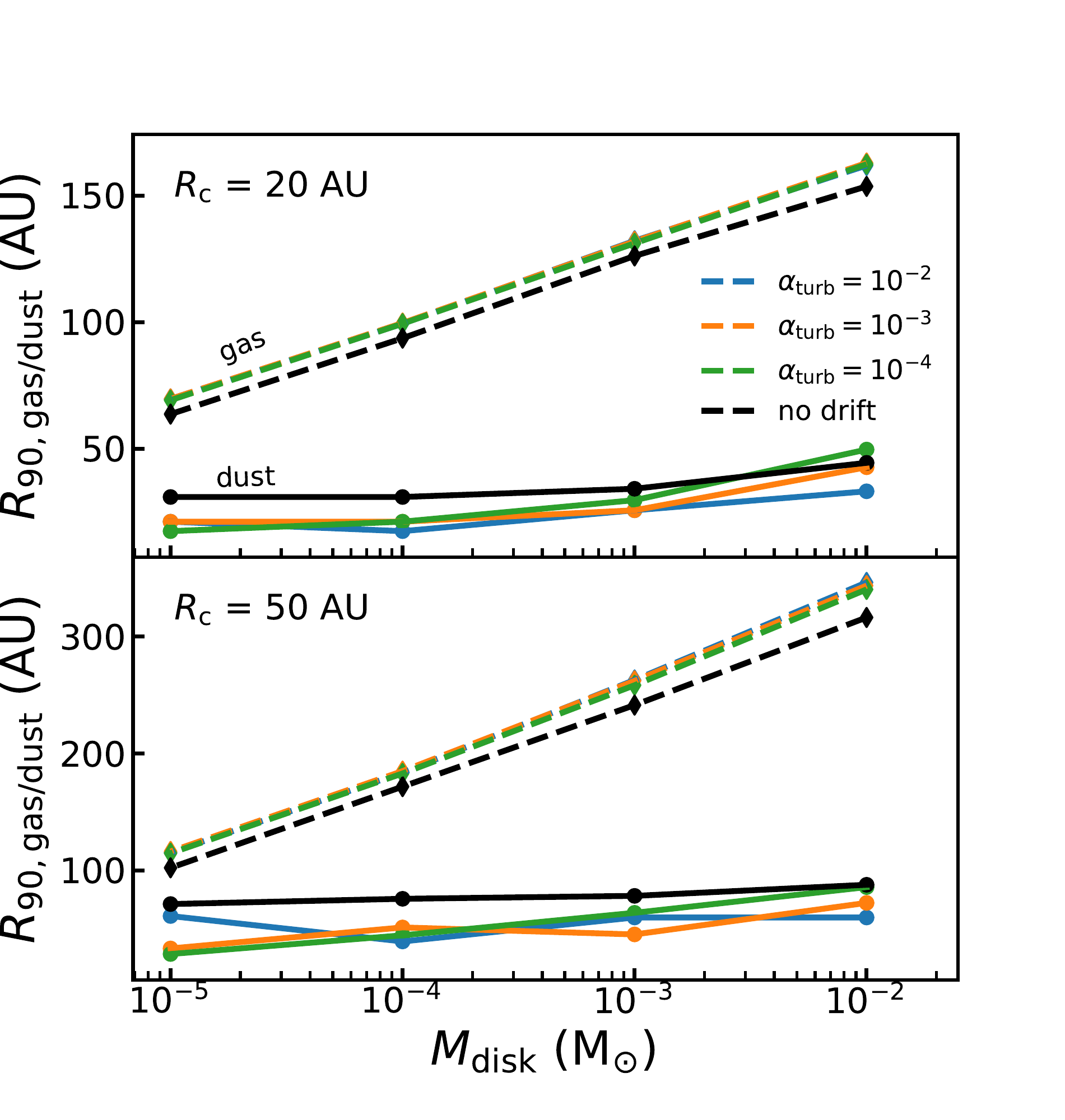}
    \caption{\label{fig: rgas and rdust versus mass} Disk outer radii versus disk mass. Top panel shows models with $\rc = 20$ AU. Bottom panel shows models with $\rc = 50$ AU. Solid lines show dust outer radii. Dashed lines show gas outer radii}
\end{figure}

The observed gas-dust size difference is also affected by the difference in optical depth. \rdust\ is calculated from the millimetre continuum emission, that is mostly optically thin, whereas \rgas\ is calculated from the optically thick \CO line emission. As optical depth is directly related to column density, the mass of the disk is also expected influence the observed outer radii.

Figure \ref{fig: rgas and rdust versus mass} shows $R_{\rm 90, gas}$ as function of \mdisk\ for our standard model as well as models with different \alp. The gas outer radii increase linearly with $\log_{10} M_{\rm disk}$, from $\sim3\times\rc$ at $\mdisk = 10^{-5}$ M$_{\odot}$ up to $\sim8\times\rc$ at $\mdisk = 10^{-2}$ M$_{\odot}$. 

This relation can be understood qualitatively using the results from Section \ref{sec: gas radial profiles}. Based on Equation \eqref{eq: rgas and rco relation} and Figure \ref{fig: rgas and rdust versus mass}, we can write $\rco \sim \rgas \propto \log_{10} \mdisk$. In the outer disk the column density scales as $N_{\rm CO} \sim \Sigma_{\rm gas} x_{\rm CO} \sim \mdisk\cdot x_{\rm CO} \cdot \exp\left(-R/\rc\right)$. At a radius \rco\ the CO column density is known and the equation can be inverted to obtain $\rco \sim \rc \log \left(\mdisk\cdot x_{\rm CO}\right)$, similar to the relation found in Figure \ref{fig: rgas and rdust versus mass}.

It should be pointed here that the dependence of \rgas\ on \mdisk\ is set by the shape of the density profile in the outer disk. In our models the density profile in the outer parts is described by an exponential, giving rise to the logarithmic dependence of \rgas\ on \mdisk. If some process is affecting the density structure of the outer disk (e.g., due to tidal truncation or external photoevaporation; \citealt{Facchini2016,Winter2018}), the relation $\rgas \propto \log \mdisk$ no longer holds. Instead the relation between \rgas\ and \mdisk\ will be set by the altered shape of the density structure in the region where $N_{\rm CO}  = 10^{15}\ \mathrm{cm}^{-2}$. 

The dust radii, shown as solid lines in Figure \ref{fig: rgas and rdust versus mass}, also increase with disk mass, but to a much smaller degree than the gas radii. Over the mass range considered here the dust radii increase by up to $1.5-2\times\rc$. This is likely due to the mm continuum emission remaining optically thin throughout most of the disk. As the disk mass increases, the total continuum flux will also increase, but the shape of the intensity profile and \rdust\ derived from it will remain largely unchanged. The small increase with dust mass can be attributed to a core of optically thick emission in the inner part of the disk. For higher disk masses this core increases in size which moves the 90\% flux contour outwards. 

Overall we find that the effect of disk mass and optical depth on \rgas\
is much larger than the effect of dust evolution on \rdust.

\subsection{\ratgasdust\ as tracer for dust evolution}
\label{sec: rgas/rdust }

\begin{figure}
    \centering
    \includegraphics[width=\columnwidth]{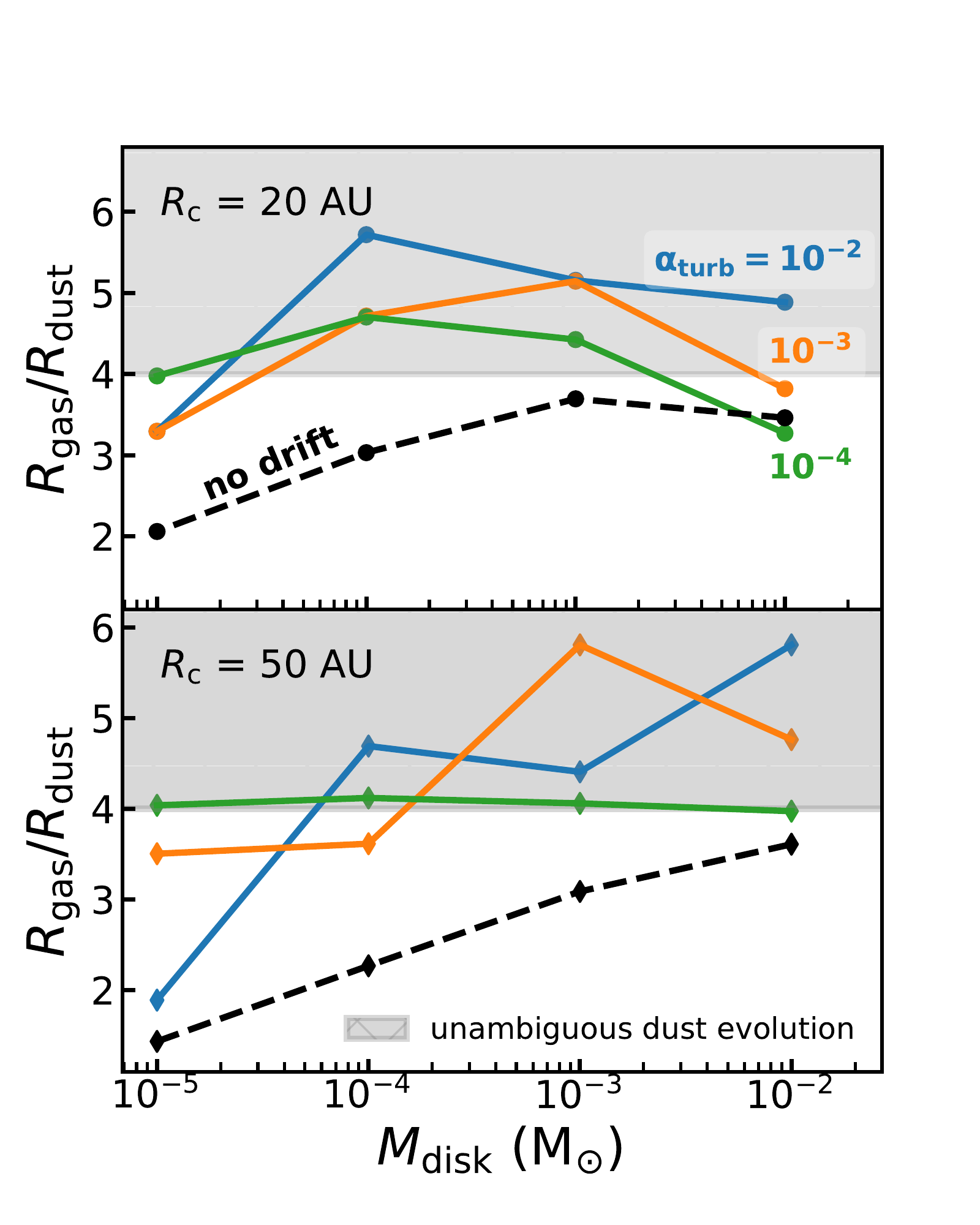}
    \caption{\label{fig: rgas over rdust versus mass} \ratgasdust\ versus disk mass. Top panel shows models with $\rc = 20$ AU. Bottom panel shows models with $\rc = 50$ AU. Dashed lines show the \textit{no drift} model, without dust evolution.}
\end{figure}

Combining the effects of dust evolution and disk mass (cf. Sections \ref{sec: alpha vs rgas and rdust} and \ref{sec: the effect of disk mass}), we now look at the gas-dust size dichotomy, quantified by \ratgasdust. Figure \ref{fig: rgas over rdust versus mass} shows \ratgasdust\ as function of \mdisk. The fiducial \textit{no drift} models show that $\ratgasdust = 1.5-3.5$, with a positive trend between \ratgasdust\ and \mdisk. Comparing to Figure \ref{fig: rgas and rdust versus mass} this trend is a direct result of \rgas\  increasing with \mdisk. 

The dust evolution models all lie above the \textit{no drift} models, indicating that for a given disk mass a model that includes radial drift and grain growth has a larger \ratgasdust\ than a model that only includes the effects of optical depth. 

For the dust evolution models with \rc\ = 50 AU, an overall positive trend of \ratgasdust\ with disk mass is found, but the trend is not monotonic and depends on \alp. For the smaller \rc\ = 20 AU models the trend becomes negative towards higher disk masses. 

From the trends in Figure \ref{fig: rgas over rdust versus mass} it is clear that the size dichotomy \ratgasdust\ can be used to identify dust evolution if the ratio is high enough ($\ratgasdust \geq 4$). Observationally, these cases are rare (see, e.g., Facchini et al., subm.). For the majority of disks, a lower ratio is observed (cf. \citealt{ansdell2018}). To identify dust evolution in these disks requires modelling of their $^{12}$CO and dust emission, taking into account their total CO and dust content. The results also show that a direct determination of \alp\ from \ratgasdust\ is not possible.

\subsection{Observational factors affecting \ratgasdust}
\label{sec: observational factors}

\subsubsection{The effect of the beamsize}
\label{sec: the effect of beamsize}

\begin{figure}
    \centering
    \includegraphics[width=\columnwidth]{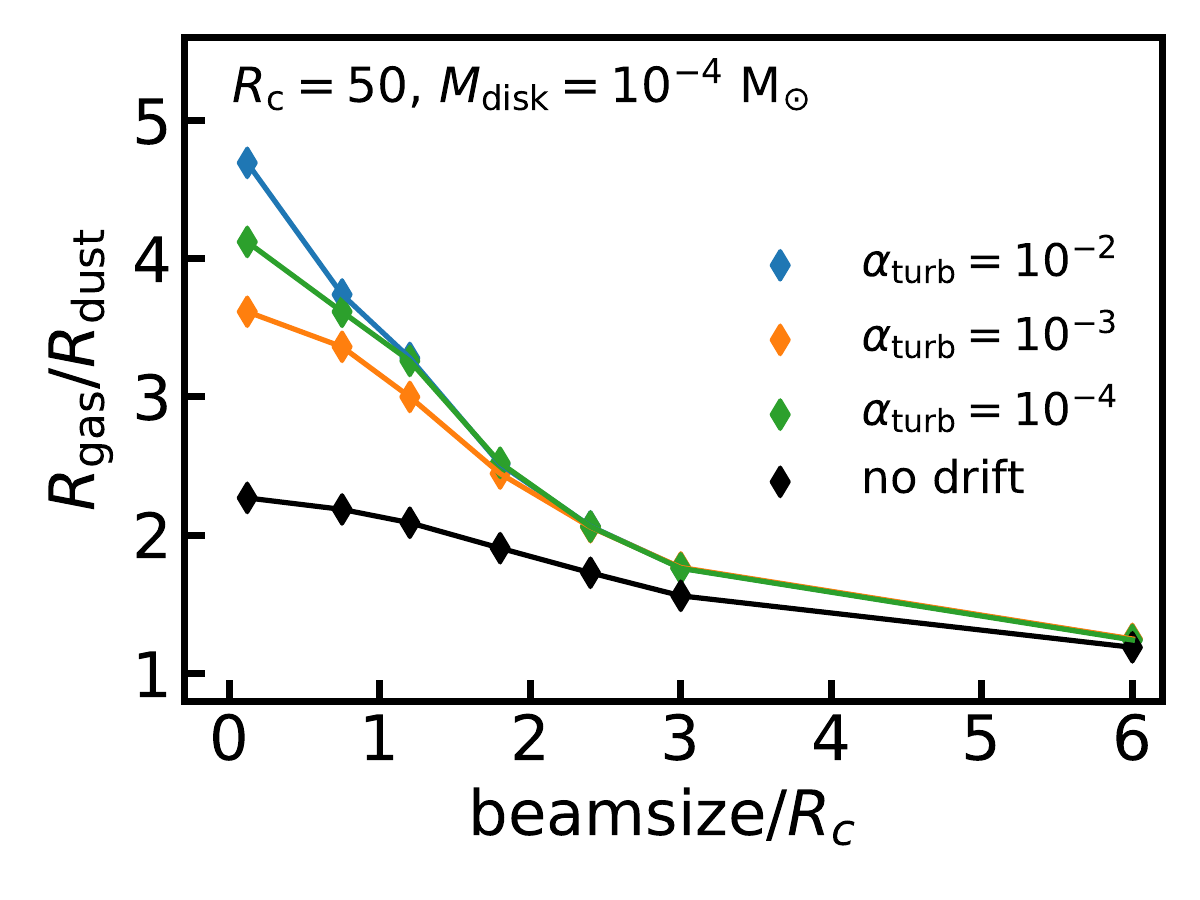}
    \caption{\label{fig: rgas over rdust versus beamsize} \ratgasdust\ versus beamsize. The effect of the beam scales with its relative size compared to the diameter of the disk. To highlight this, the beamsize is expressed in terms of the characteristic size of the disk. Similar figures for different \mdisk\ and \rc\ are shown in Figure \ref{fig: rgas over rdust versus beamsize onepage}}
\end{figure}

Observational factors such as the size of the beam and the background noise level are also able to influence the gas-dust size dichotomy. Convolution with the beam smears out the intensity profile. For a centrally peaked intensity profile this will move $R_{\rm 90}$ outward. For a beamsize (FWHM$_{\rm beam}$) much larger than the observed disk the intrinsic differences between the gas and dust emission are washed out and \ratgasdust\ is expected to approach unity. 

Figure \ref{fig: rgas over rdust versus beamsize} shows the effect of beamsize on \ratgasdust\ for an example disk with $\mdisk = 10^{-4}\ \mathrm{M}_{\odot}$ and $\rc = 50$ AU. Similar panels for the other models are shown in Figure \ref{fig: rgas over rdust versus beamsize onepage}. \ratgasdust\ decreases with beamsize, approaching unity when the beamsize becomes $\sim3\times \rc$. At a beamsize$\sim1\times\rc$, \ratgasdust\ has dropped below 4 and dust evolution can no longer be unambiguously be identified using only \ratgasdust\ (see Section \ref{sec: rgas/rdust }). However, if the uncertainties on \ratgasdust\ are sufficiently small and the total CO content of the disk is known, dust evolution can still be inferred from observations with FWHM$_{\rm beam} \leq 2 \rc$. 

\subsubsection{The effect of noise level}
\label{sec: the effect of SNR}
\begin{figure}
    \centering
    \includegraphics[width=\columnwidth]{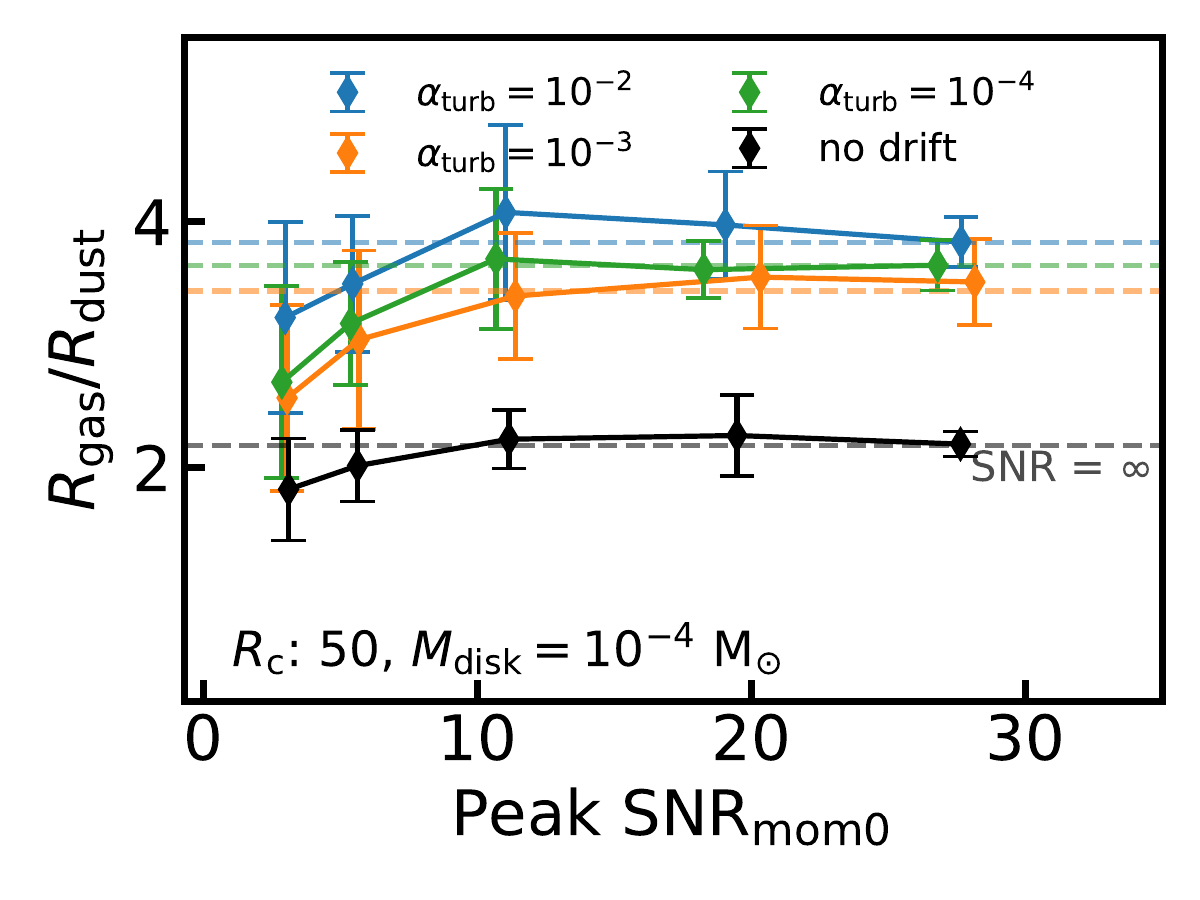}
    \caption{\label{fig: rgas over rdust versus SNR} \ratgasdust\ versus peak SNR in the moment 0 map of the $^{12}$CO emission. Similar figures for different \mdisk\ and \rc\ are shown in Figure \ref{fig: rgas over rdust versus SNR onepage} }
\end{figure}

Noise has two ways in which it can interact with the observational outer radius. It affects the shape of the curve-of-growth, thus changing the radius that encloses 90\% of the total flux. In addition, the noise in the image sets the uncertainty on the total flux, which propagates through into the errors on $R_{\rm 90}$. For the gas-dust size difference, the noise on the gas emission is dominant. This is due to a difference in bandwidth: the gas emission is narrow in frequency, typically $\sim 0.5 \mathrm{km\ s}^{-1}$, whereas the continuum emission uses the full bandwidth of the observations. For example, a typical ALMA band 6 observation that has three continuum spectral windows (with a total bandwidth $\Delta\nu_{\rm bandwidth} \simeq 2$ GHz) and 1 line spectral window centred on the $^{12}$CO line (with line width $\Delta\nu_{\rm line} \simeq 0.384$ MHz) the noise on the continuum is a factor $\sqrt{\Delta\nu_{\rm bandwidth}/\Delta\nu_{\rm line}}\sim50$ lower compared to the line emission.

To simulate the effect of noise, empty channels taken from ALMA observations are added to the model image cube after convolution. The size of the convolution beam used for the model was matched to that of observations of Lupus disks ($0".25$; \citealt{ansdell2018}). The rms of the noise was scaled to obtain the requested peak SNR in the moment 0 map for the models.

The results are shown in Figure \ref{fig: rgas over rdust versus SNR} for the same example disk ($\mdisk = 10^{-4}\ \mathrm{M}_{\odot},\ \rc = 50\ \mathrm{AU}$). The other disks are shown in Figure \ref{fig: rgas over rdust versus SNR onepage}. The average \ratgasdust\ measured at low SNR$_{\rm mom 0}$ is smaller than the noiseless case. As the SNR$_{\rm mom 0}$ increases it converges to the value measured in the absence of noise. A peak SNR$_{\rm mom 0}$ $\sim 10$ is sufficient to recover the \ratgasdust\ of the noiseless case. The uncertainties on \ratgasdust\ are reduced when the peak SNR$_{\rm mom 0}$ increases, down to $\leq10\%$ at a peak SNR$_{\rm mom 0}$ of $\sim30$. Note however that the noiseless \ratgasdust\ is recovered within the errorbars of the measured \ratgasdust\ already at peak SNR$_{\rm mom 0}$ $\sim 5$. 

Summarising the effect of the observational factors, two recommendations for future observations can be made. Firstly, differentiating between only optical depth and dust evolution in addition to optical depth requires FWHM$_{\rm beam} \leq 1\times \rc$. For a disk with $\rc = 20$ at 150 pc, this means a beamsize of $0\farcs14 = 20$ AU. Secondly, to accurately measure the gas-dust size difference requires a peak SNR $\geq 10$ in the $^{12}$CO moment zero. Note that an increased sensitivity will improve the uncertainty on the measured \ratgasdust\ and can thus better distinguish cases where \ratgasdust\ is unambiguously $> 4$. 

\section{Discussion}
\label{sec: discussion}

\subsection{CO underabundance and \rgas}
\label{sec: CO depletion}

In our models we have assumed standard ISM abundances for carbon and oxygen, resulting in an overall CO abundance of $x_{\rm CO} \sim 10^{-4}$. However, recent observations have found CO to be underabundant by a factor $10-100$ with respect to the ISM in several disks (e.g., \citealt{ Favre2013,Kama2016,Cleeves2016,McClure2016}). The low CO-based disk gas masses found by recent surveys suggest that CO could be underabundant in most disks (see, e.g., \citealt{ansdell2016,miotello2017,Long2017}). As shown in Section \ref{sec: the effect of disk mass} \rgas\ is directly related to the CO content of the disk. The observed underabundance of CO in disks will result in them having a smaller observed gas disk size compared to our models. 

\begin{figure}
    \centering
    \includegraphics[width=\columnwidth]{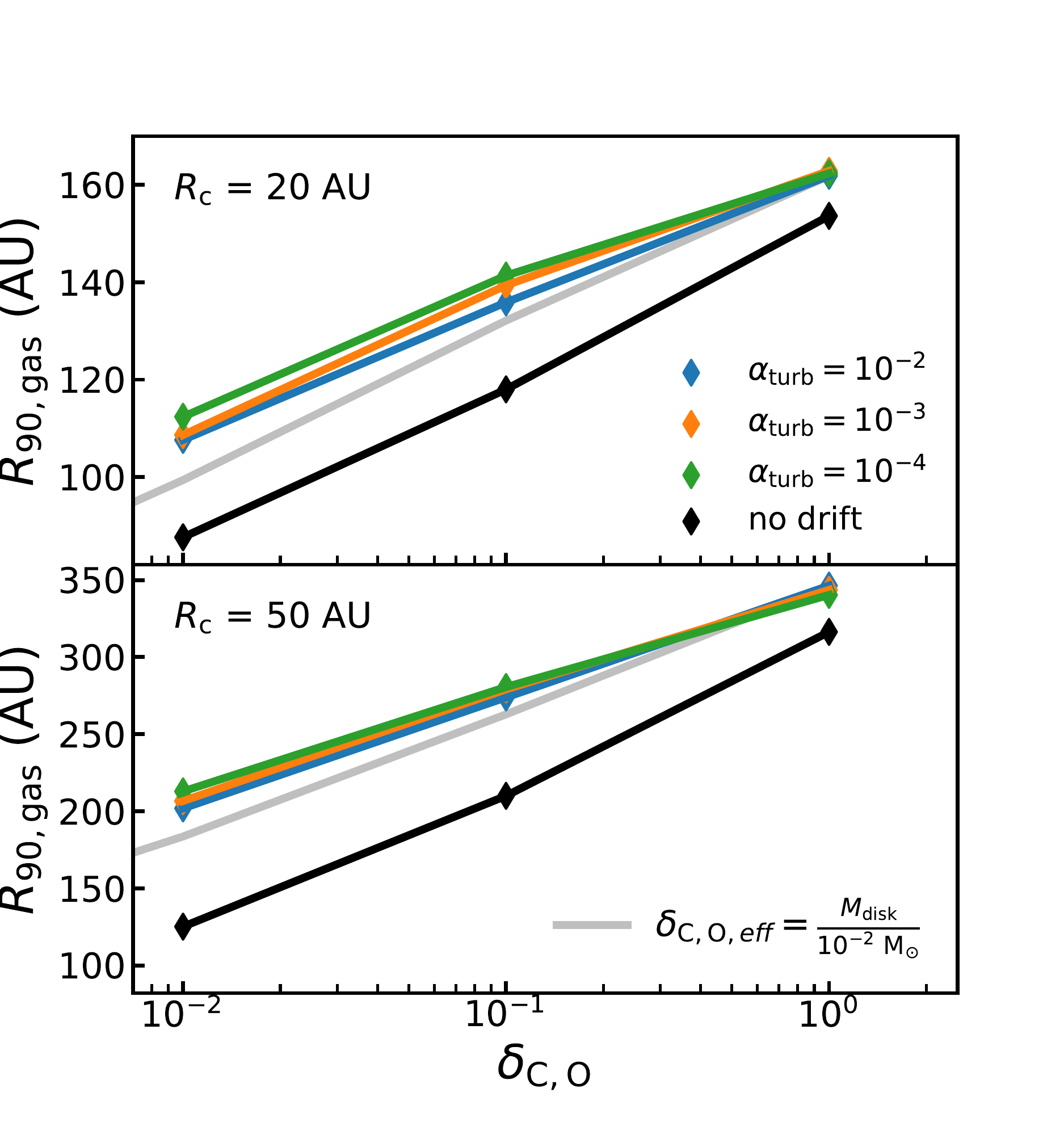}
    \caption{\label{fig: rgas and rdust versus CO underabundance} Disk outer radii versus CO underabundance. For comparison, the outer radii for the \textit{no drift} model is shown in gray. Top panel shows models with $\rc = 20$ AU. Bottom panel shows models with $\rc = 50$ AU. }
\end{figure}

To quantify the effect of CO underabundance on the measured \rgas\ the models with $\mdisk = 10^{-2}\ \mathrm{M}_{\odot}$ were rerun, but now the total amount of carbon and oxygen in the disk is reduced by a factor $\delta_{C,O} = 0.1-0.01$, mimicking the observed underabundance of CO. Figure \ref{fig: rgas and rdust versus CO underabundance} shows that \rgas\ decreases linearly with $\log_{10} \delta_{C,O}$, similar to the $\rgas \propto \log_{10} \mdisk$ found earlier. This highlights again the importance of total CO content of the disk: \rgas\ measured from a disk with higher mass but underabundant in CO (e.g., \mdisk, \dCO) = ($10^{-2}$ M$_{\odot}$, 0.1) is very similar to the observed \rgas\ of a disk with standard ISM abundances but that has lower disk gas mass (e.g., \mdisk, \dCO) = ($10^{-3}$ M$_{\odot}$, 1), because both of them have a similar CO content.

\begin{figure}
    \centering
    \includegraphics[width=\columnwidth]{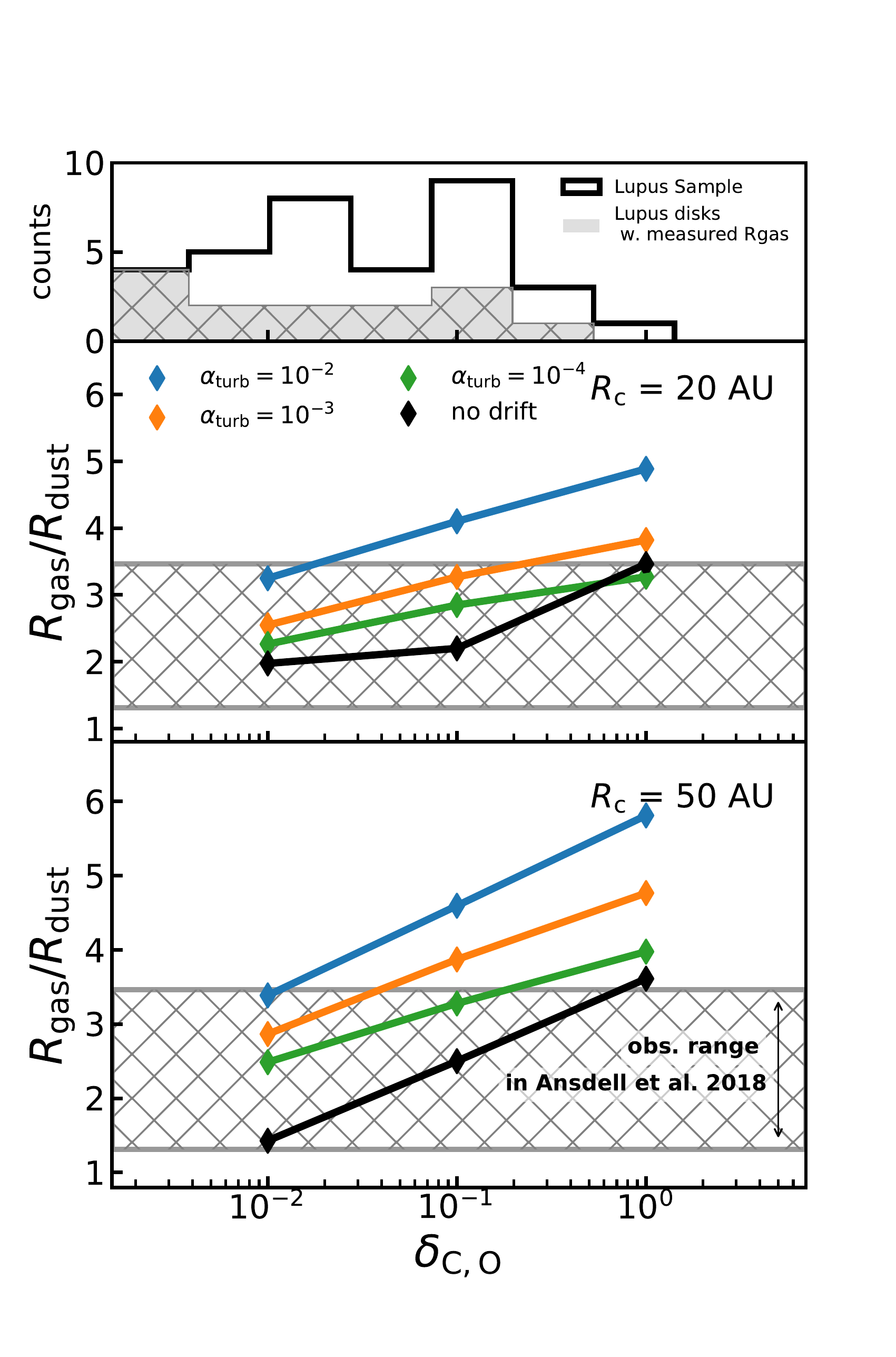}
    \caption{\label{fig: rgas over rdust versus CO underabundance} \ratgasdust\ versus CO underabundance. Middle panel shows models with $\rc = 20$ AU. Bottom panel shows models with $\rc = 50$ AU. The observed range of \ratgasdust\ from \cite{ansdell2018} is shown in gray. Top panel shows a histogram of the gas-to-dust ratios measured in Lupus \citep{ansdell2016,miotello2017}. These have been converted into an effective CO underabundance using $\delta_{\rm C,O, eff} = \Delta_{\rm gd}/100$, where \gdratio\ is the gas-to-dust mass ratio.}
\end{figure}

A lower \rgas\ will also result in a lower \ratgasdust. Figure \ref{fig: rgas over rdust versus CO underabundance} shows \ratgasdust\ as function of CO underabundance $\delta_{\rm C,O}$. In the case of no CO underabundance ($\delta_{C,O} = 1$), $\ratgasdust = 2.5-5.5$. By decreasing the amount of CO in the disk by a factor of 100, the gas disk size decreases leading to $\ratgasdust = 1.5$ for the \textit{no drift} model and $\ratgasdust = 2-3.5$ for the dust evolution models. Note that these values are for a disk with $M_{\rm disk} = 10^{-2}\ \mathrm{M}_{\odot}$. For a less massive disk that is also underabundant in CO \ratgasdust\ will be lower (cf. Section \ref{sec: the effect of disk mass}). 

For a sample of 22 disks \cite{ansdell2018} measured \rgas\ from the $^{12}$CO 2-1 emission and \rdust\ from the 1.3 mm continuum emission and found $\ratgasdust = 1.5-3.5$. The sample is skewed towards the most massive disks in their sample, with $\mdust = 0.5-2.7 \times 10^{-4}\ \mathrm{M}_{\odot}$. This makes them comparable in dust content to the models discussed in this section ($\mdust = 10^{-4}\ \mathrm{M}_{\odot}$). A simple, first order comparison, between the models and the observations can therefore be made.
CO underabundances of the disks in the Lupus sample are calculated by assuming a gas-to-dust mass ratio of 100 and comparing that to the ratio of $M_{\rm CO\ based}/\mdust$, where $M_{\rm CO\ based}$ is the CO-based gas mass estimate from \cite{miotello2017}. For example, a disk with $M_{\rm CO\ based} = 10^{-3}\ \mathrm{M}_{\odot}$ and $\mdust = 10^{-4}\ \mathrm{M}_{\odot}$ is interpreted as having a CO underabundance of $\delta_{\rm C,O} = \tfrac{M_{\rm CO\ based}}{\mdust}/100 = 0.1$. 

The calculated CO abundances show that several disks in the sample have $\delta_{C,O} \leq 10^{-2}$. For the same CO underabundance, the \textit{no drift} model has $\ratgasdust = 1.5-2.0$, which is at the low end or below the observed range, suggesting that a least for some of the sources in the sample explaining the observed \ratgasdust\ requires dust evolution. Modelling of the individual sources is required to provide a definitive identification of dust evolution, which is beyond the scope of this work (but see Trapman et al. subm.). 

\subsection{The effect of the surface density slope on outer radii}
\label{sec: the effect of gamma}

\begin{figure}
    \centering
    \includegraphics[width=\columnwidth]{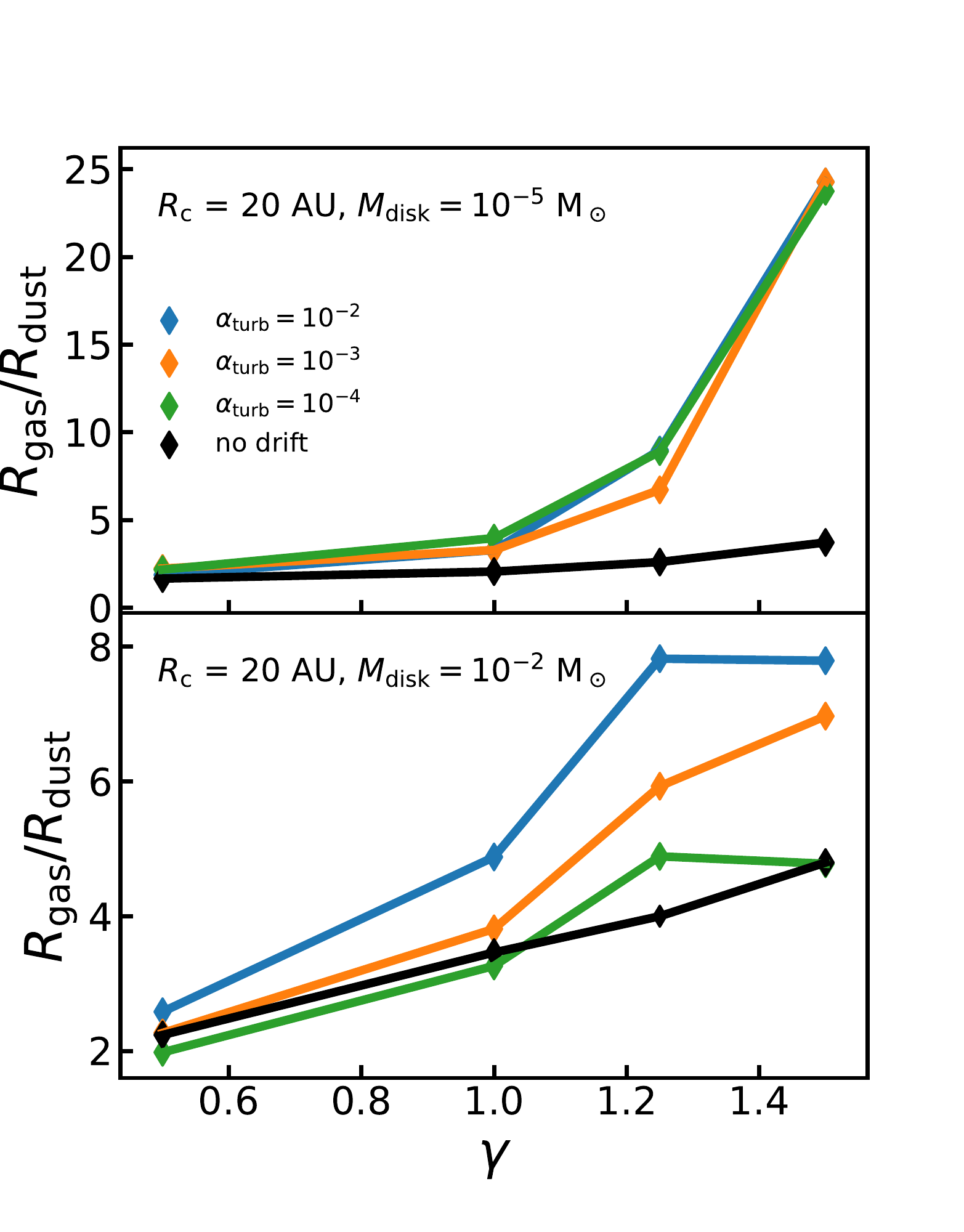}
    \caption{\label{fig: gamma} \ratgasdust\ as function of the slope of the gas surface density. Top panel shows models with $\mdisk = 10^{-5}\ \mathrm{M}_{\odot}$. Bottom panel shows models with $\mdisk = 10^{-2}\ \mathrm{M}_{\odot}$.}
\end{figure}

The surface density is governed by three parameters: \rc, \mdisk\ and $\gamma$ (cf. Eq. \ref{eq: tapered powerlaw}). The slope $\gamma$ sets how the material is distributed in the disk and therefore affects the outer radius of the disk. In addition, the physical processes involved in dust evolution are also affected by $\gamma$. 
The slope of the gas surface density is not well constrained, having been observational constrainted only for a few disks (e.g., \citealt{Cleeves2016,WilliamsMcPortland2016,Zhang2017,Miotello2018}). Most of these studies find a value of $\gamma \sim 1.0$. In this section we investigate how much \ratgasdust\ depends on $\gamma$. Figure \ref{fig: gamma} shows \ratgasdust\ versus $\gamma = [0.5,1.0,1.5]$ for a set of low mass and high mass models. For low mass disks ($\mdisk = 10^{-5}\ \mathrm{M}_{\odot}$), \ratgasdust\ increases drastically when $\gamma$ is increased from 1.0 to 1.5, with values of $\ratgasdust \simeq 24$ for models with dust evolution.  

For the disks with dust evolution and $\gamma = 1.5$, mm-sized grains have been removed from the disk except for the inner few AU. As a result, the continuum emission is concentrated in this inner region and a very small dust outer radius is inferred. For the \textit{no drift} model the dust radius is not similarly affected and the gas-dust size difference only increases to $\ratgasdust = 3.7$.  

Increasing $\gamma$ for the high mass disk does not have a significant effect on \ratgasdust.  For the high mass disk the trends are similar for the \textit{no drift} model and the dust evolution models. Here dust evolution is not significantly affected by the change in $\gamma$. For both low and high mass disks, decreasing $\gamma$ from 1.0 to 0.5 results in a \ratgasdust\ that is lower by a factor $\sim 2$.

\subsection{How well does the observed R$_{\rm out}$ match to physical size of the disk}
\label{sec: mass fraction vs flux fraction}

\begin{figure}
    \centering
    \begin{subfigure}{0.99\columnwidth}
    \includegraphics[width=\columnwidth]{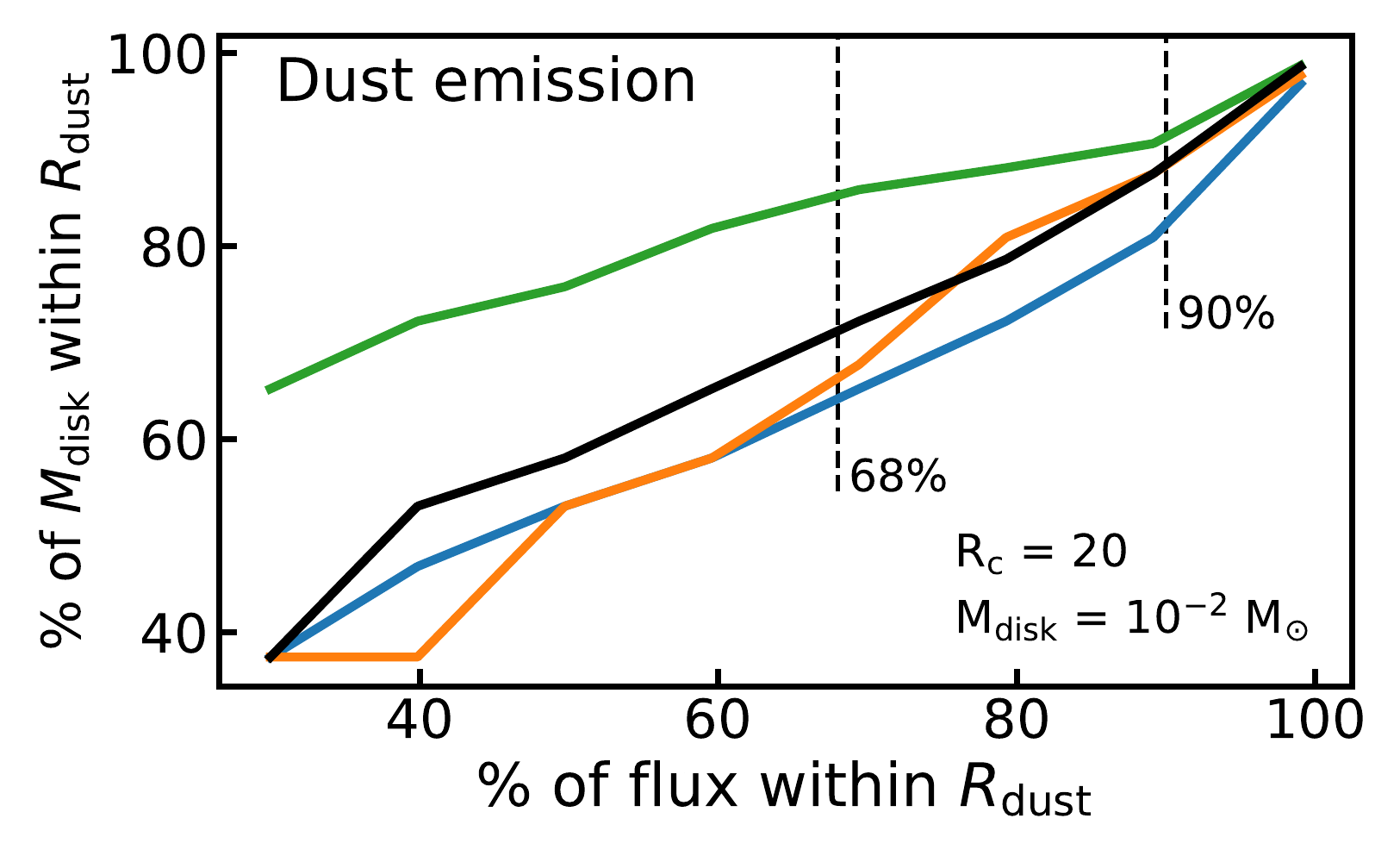}
    \end{subfigure}
    \qquad
    \begin{subfigure}{0.99\columnwidth}
    \includegraphics[width=\columnwidth]{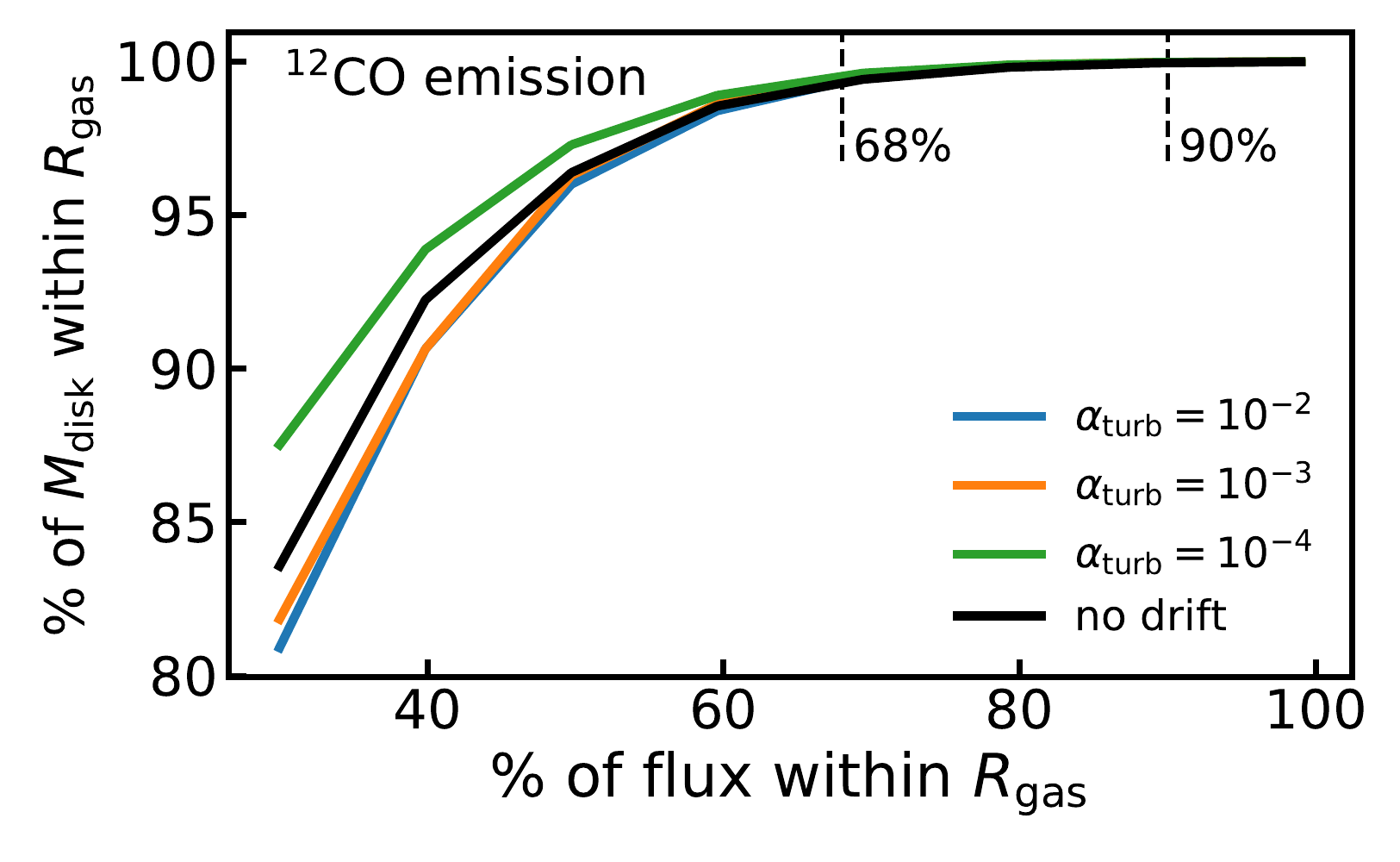}
    \end{subfigure}
    \caption{\label{fig: mass frac vs flux frac} Fraction of flux $f$ used to calculate $R_{\rm dust}$ and $R_{\rm 90, gas}$ compared to the fraction of \mdisk\ within $R_{\rm dust}$ and $R_{\rm 90, gas}$. Top panel shows the dust emission and the bottom panel shows the gas emission. }
\end{figure}

In this work we have quantified the radial extent of the disk using $R_{\rm 90}$, a flux based measure of the size of the disk. For this radius to be the outer edge of the disk in a physical sense, one could require that it encloses most (e.g., $\geq 90\%$) of the total mass of the disk. 

In the cumulative intensity method used to define \rgas\ and \rdust, a free parameter is the fraction of total flux $f$ used (cf. Eq. \ref{eq: outer radius}). In this work the outer radius is set at 90\% of the total flux under the assumption that this radius will enclose most of the disk. Here we investigate the requirement on $f$ if we want the outer radius to enclose $\geq 90\%$ of the mass.

Figure \ref{fig: mass frac vs flux frac} shows, for a given $f$ used to compute $R_{\rm 90, gas}$ and $R_{\rm 90, dust}$, what fraction of the total disk mass is enclosed within this outer radius. Similar figures for the other disks in the model grid are shown in Appendix \ref{app: mass fractions}. 

For the dust emission, show in the top panel of Figure \ref{fig: mass frac vs flux frac}, the fraction of enclosed disk mass ($f_{\rm mass}$) correlates with the fraction of total flux ($f_{\rm flux}$) used to define the outer radius. The relation between $f_{\rm mass}$ and $f_{\rm flux}$ is roughly linear, however the exact trend depends on \alp, \mdisk\ and \rc. Using a flux fraction of $f_{\rm flux}=0.9$, between 75\% and 90\% of the total mass is enclosed with the exact fraction depending on \alp. At $f_{\rm flux}=0.68$ this has dropped to between 60\% and 84\% of the total mass. 

The bottom panel of Figure \ref{fig: mass frac vs flux frac} shows that for the $^{12}$CO emission, any gas outer radius $R_{\rm 90, gas}$ defined using a fraction of total flux $f_{\rm flux} > 60\%$ encloses almost all ($>98\%$) of the total disk mass and would thus meet our criterion of a physical outer radius (i.e., enclosing $\geq 90\%$ of \mdisk). Note that by the definition of eq. \eqref{eq: outer radius}, these observational outer radii are not the same size. For example, a radius enclosing 90\% of the flux has to be larger than a radius enclosing 75\% of the flux. These observational outer radii are all related through Equation \eqref{eq: rgas and rco relation} to the same physical point in the disk, where the CO column density equals $10^{15}\ \mathrm{cm}^{-2}$ (cf. Section \ref{sec: gas radial profiles}).

Summarising, the fraction of mass enclosed by $R_{\rm dust}$ scales roughly linearly with the fraction of continuum flux used to define $R_{\rm dust}$. To have the dust radius enclose most of the disk mass, the outer radius should be defined using a high fraction ($90-95\%$) of the total flux. For the gas, any radius enclosing $> 60\%$ of the flux will contain most of the mass.

\section{Conclusions}
\label{sec: conclusions}
The gas in protoplanetary disks is found to be universally more extended than the dust. This effect can result from grain growth and subsequent inward drift of mm-sized grains. However, the difference in line optical depth between the optically thick $^{12}$CO emission of the gas and the optically thin continuum emission of the dust also produces a gas-dust size dichotomy. In this work the thermochemical code DALI \citep{Bruderer2012,Bruderer2013}, extended to include dust evolution \citep{Facchini2017}, is used to run a grid of models. Using these models, the impact of dust evolution, optical depth and disk structure parameters on the observed gas-dust size difference are investigated. Our main conclusions can be summarised as follows:

\begin{itemize}
    \item Including dust evolution leads to smaller observationally derived dust radii and larger gas radii. Dust evolution, as quantified by \alp, has a complex effect on the dust radius that also depends on the disk mass and the characteristic radius. The gas outer radius is unaffected by changes in \alp.
    \item The gas outer radius \rgas\ is directly related to the radius at which the CO column density drops below $10^{15}\ \mathrm{cm}^{-2}$ where CO becomes photodissociated. \rgas\ scales with the product  $\mdisk\cdot x_{\rm CO}$, the total CO content of the disk. $R_{\rm gas}$ is directly related to the radius where $^{12}$CO no longer is able to self-shield.
    \item \ratgasdust\ increases with the total CO content and is higher for disks that include dust evolution. Disks with $\ratgasdust > 4$ are difficult to explain without dust evolution. For $\ratgasdust < 4$, deducing whether or not a disk is affected by dust evolution from the size ratio requires a measure of the total CO content. However, constraining \alp\ using \ratgasdust\ is not possible.
    \item Increasing the beamsize and lowering the peak SNR of the $^{12}$CO moment 0 map both decrease the measured \ratgasdust. To minimize the effect of these observational factors requires FWHM$_{\rm beam} \leq 1\times \rc$ and SNR$_{\rm peak, mom 0} > 10$.
    \item \ratgasdust\ increases with the slope of the surface density $\gamma$. In low mass disks with high $\gamma$, dust evolution removes almost all grains from the disk, resulting in large gas-dust size differences ($\ratgasdust \sim 24$).
    \item To have the dust radius enclose most of the disk mass, the outer radius should be defined using a high fraction ($90-95\%$) of the total flux. For the gas, any radius enclosing $> 60\%$ of the flux will contain most of the mass. 
\end{itemize}

The gas-dust size dichotomy is predominantly set by the structure and CO gas content of the disk, which can produce size differences up to $\ratgasdust \sim 4$. Disks with $\ratgasdust > 4$ can be directly identified as having undergone dust evolution, provided the gas and dust radii were measured with FWHM$_{\rm beam} \leq 1\times \rc$. However, these disks are rare in current observations. For disks with a smaller gas-dust size difference, modelling of the disk structure including the total CO gas content is required to identify radial drift and grain growth. 

\begin{acknowledgements}
We would like to thank Dr. G. Rosotti for the useful discussions and we thank the anonymous referee for the useful comments that helped improve the paper.
LT and MRH are supported by NWO grant 614.001.352. Astrochemistry in Leiden is supported by the Netherlands Research School for Astronomy (NOVA). SF is supported by an ESO fellowship. All figures were generated with the \texttt{PYTHON}-based package \texttt{MATPLOTLIB} \citep{Hunter2007}.
\end{acknowledgements}

\bibliographystyle{aa} 
\bibliography{references}


\begin{appendix}
\section{The effect of inclination}
\label{app: inclination}

\begin{figure}
    \centering
    \begin{subfigure}{0.99\columnwidth}
    \includegraphics[width=\columnwidth]{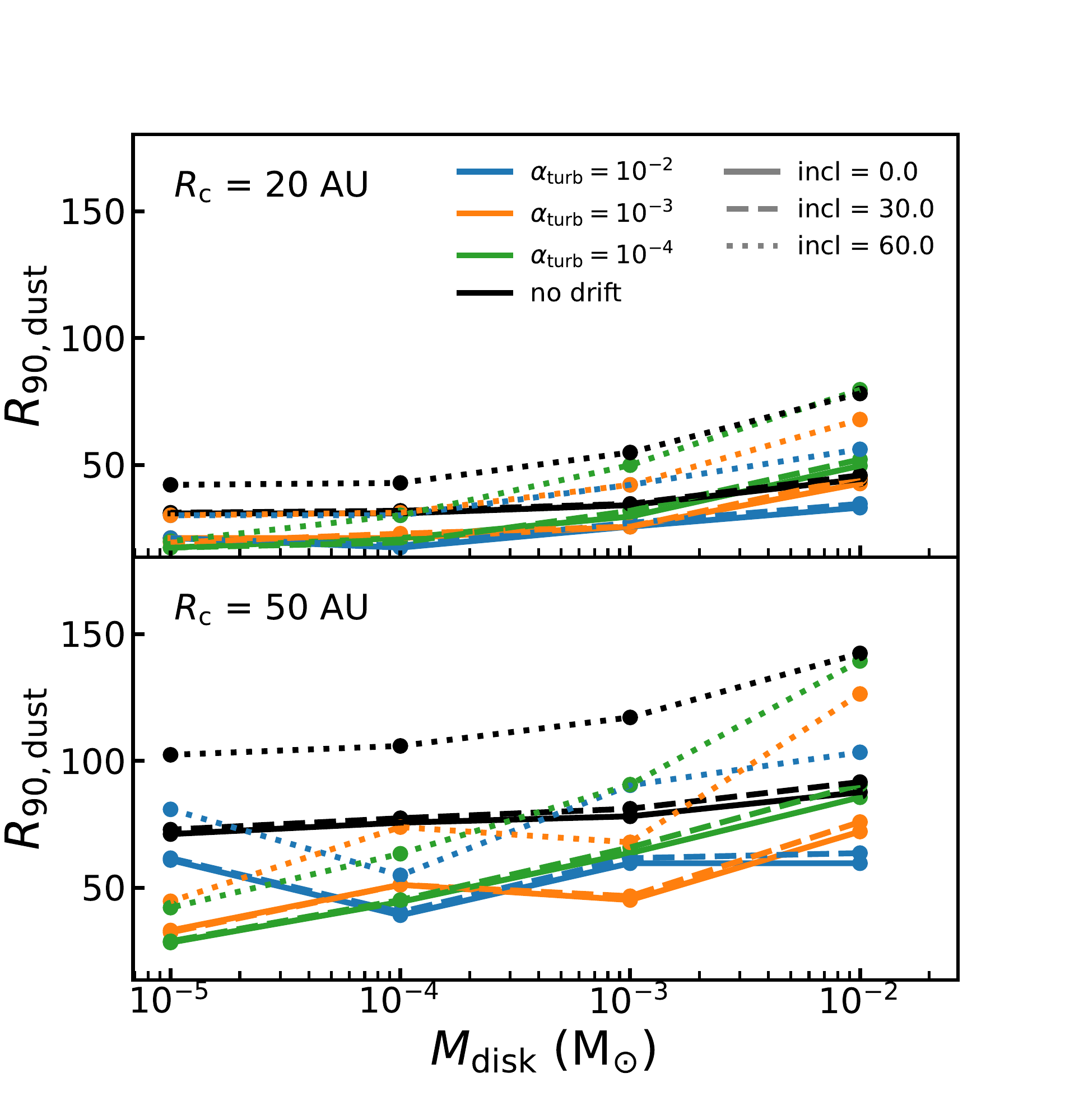}
    \end{subfigure}
    \caption{\label{fig: inclination dust} The effect of inclination on measuring the dust outer radius. Disks with a $R_C = 20$ AU and $R_C = 50$ AU are shown in the top and bottom panel, respectively. }
\end{figure}

\begin{figure}
    \centering
    \begin{subfigure}{0.99\columnwidth}
    \includegraphics[width=\columnwidth]{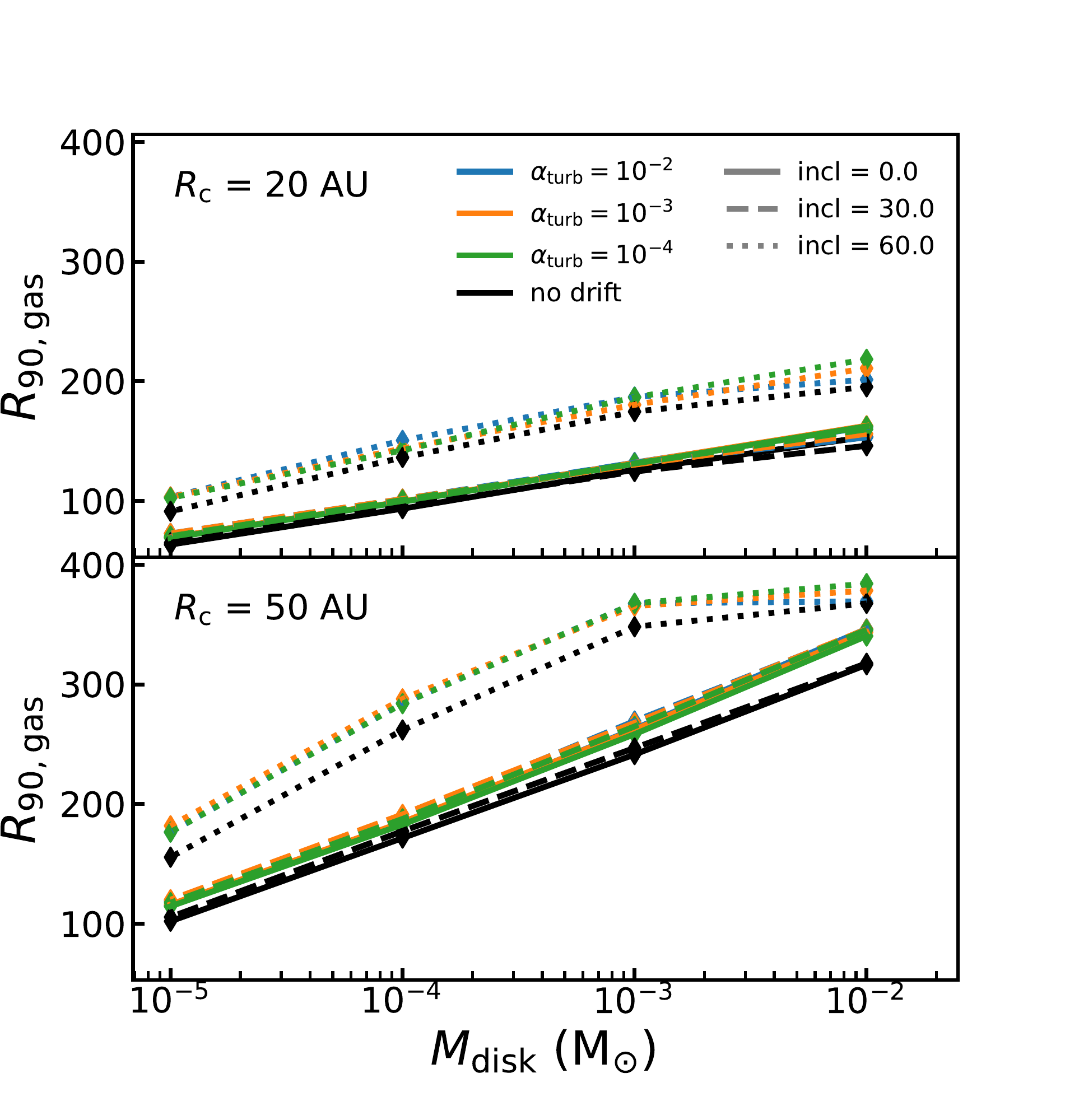}
    \end{subfigure}
    \caption{\label{fig: inclination gas} The effect of inclination on measuring the gas outer radius. Disks with a $R_C = 20$ AU and $R_C = 50$ AU are shown in the top and bottom panel, respectively. }
\end{figure}

In the analysis in this work all disk radii were measured from disks with an inclination of 0 degrees. Inclination increases the optical depth along the line of sight. This can affect the measured size of the disk, especially for the gas, which is determined from optically thick emission (cf. Section \ref{sec: measuring the outer radius}). 

Figures \ref{fig: inclination dust} and \ref{fig: inclination gas} shows gas and dust radii measured from images with an inclination of $i=0,30,60$ degrees. For the inclined images, the cumulative flux is calculated using elliptical apertures instead to account for the projection. Between $i=0^{\circ}$ and $i=30^{\circ}$ there is no noticeable difference in the outer radii. For 60 degrees outer radii have come slightly larger, but even at the most extreme the effect is smaller than a 50\% increase. Thus only for disks with high inclination ($i>60^{\circ}$) should the effect of inclination be considered when trying to identify dust evolution. 
\section{Measuring \rgas\ from $^{13}$CO 2-1 moment zero maps}
\label{app: 13CO outer radii}

\begin{figure}
    \centering
    \includegraphics[width=\columnwidth]{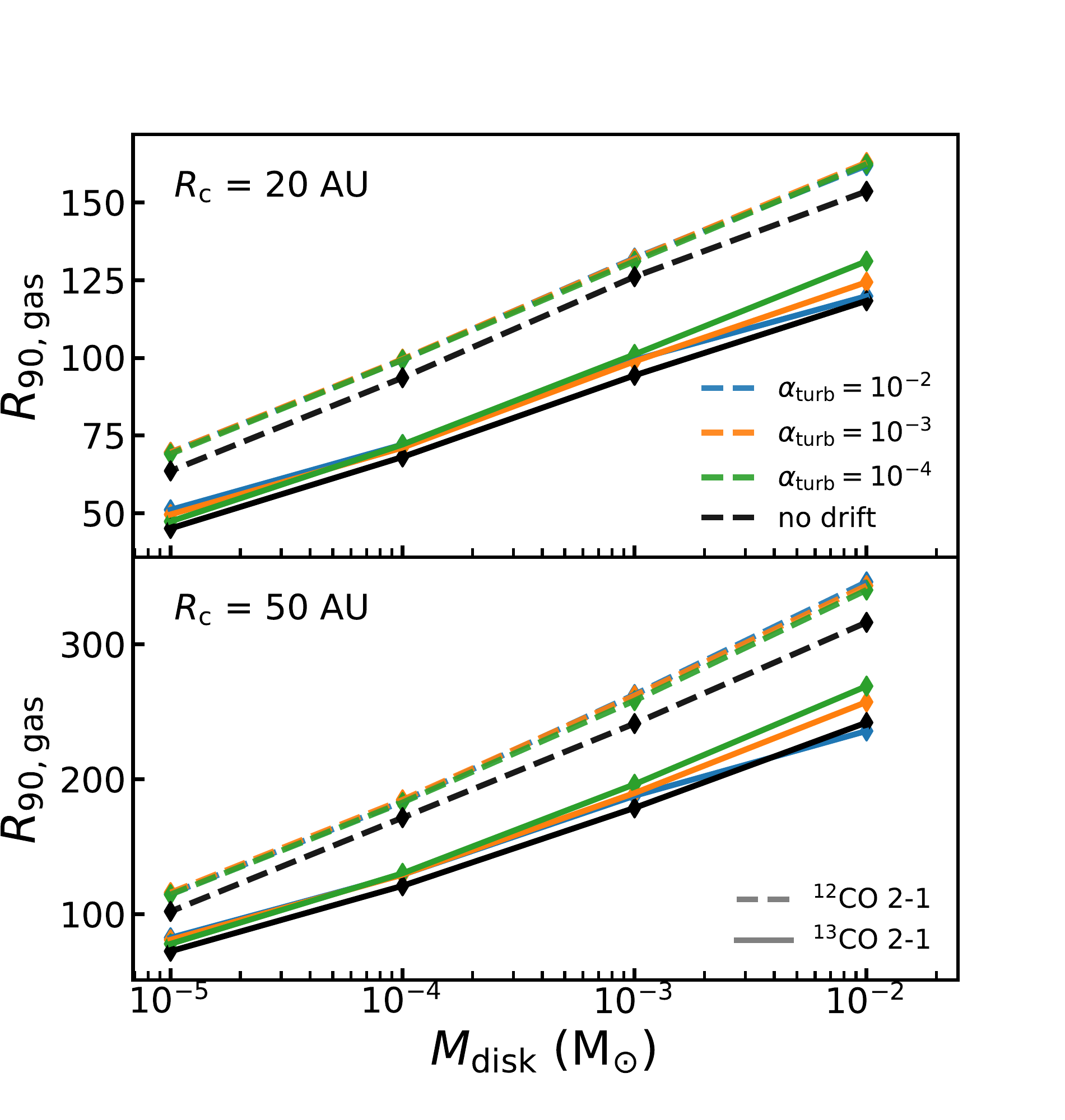}
    \caption{\label{fig: 13CO outer radii} Comparison of gas outer radii measured from $^{13}$CO 2-1 emission (solid lines) and $^{12}$CO 2-1 emission (dashed lines). }
\end{figure}

\begin{figure}
    \centering
    \includegraphics[width=\columnwidth]{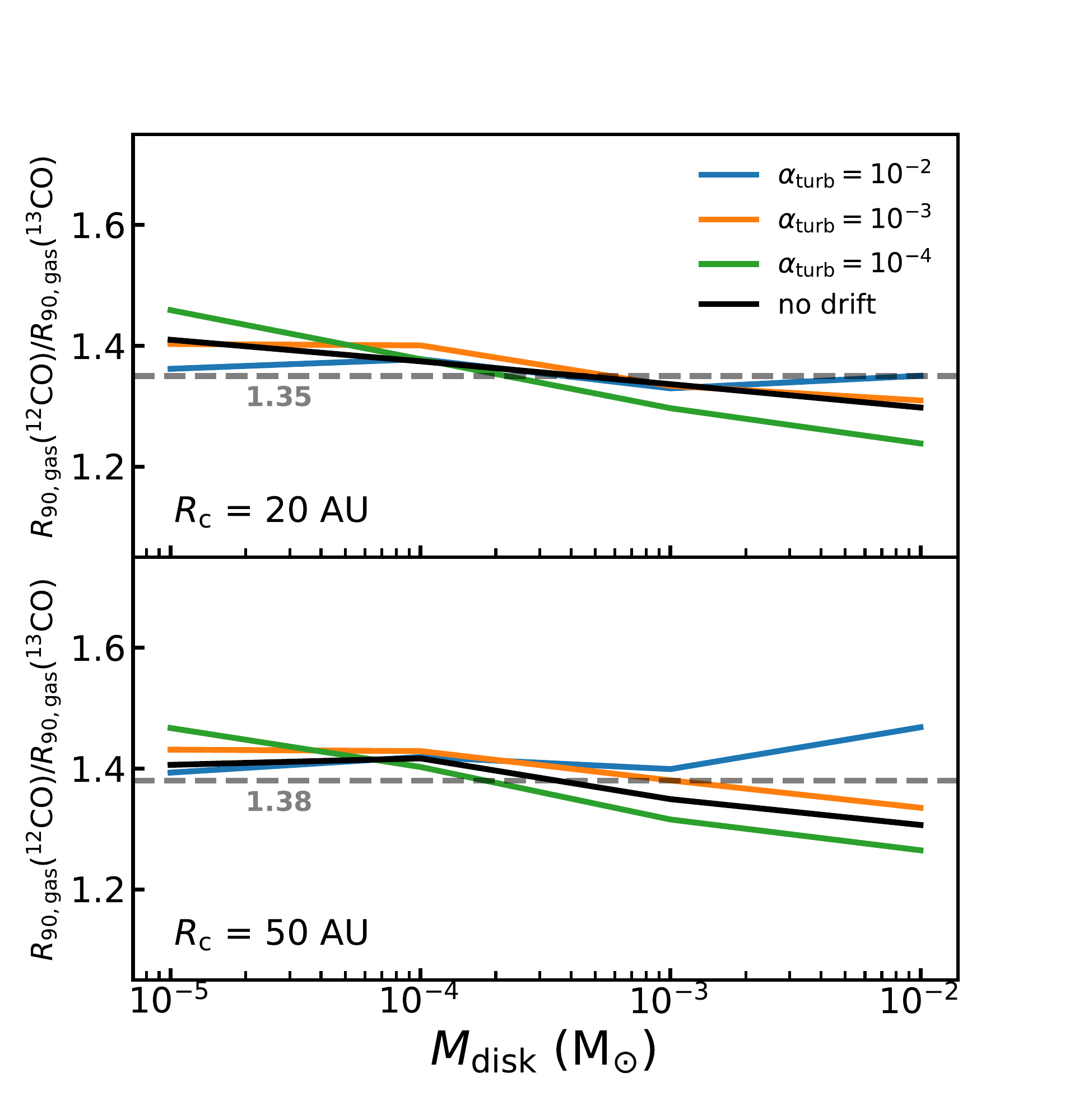}
    \caption{\label{fig: 13CO outer radii ratio} Ratio of gas outer radii measured from $^{12}$CO 2-1 emission (\rgas($^{12}$CO)) over $^{13}$CO 2-1 emission (\rgas($^{13}$CO)). The mean ratio of all models is shown in gray. }
\end{figure}

Here we investigate how the gas outer radius would differ, if instead it had been defined as the radius enclosing 90\% of the $^{13}$CO 3-2 flux. $^{13}$CO is added to the model parametrically, by taking the CO abundances and scaling them with the $^{12}$C/$^{13}$C elemental ratio, assumed to be $^{12}$C/$^{13}$C $= 77$. 

Figure \ref{fig: 13CO outer radii} shows \rgas($^{13}$CO) and \rgas($^{12}$CO) as function of disk mass.\rgas($^{13}$CO) also with disk mass in a similar manner to \rgas($^{12}$CO), i.e., \rgas($^{13}$CO) $\sim \log_{10} M_{\rm disk}$. As a result, using the $^{13}$CO emission will not change the qualitative results seen in this work.

Figure \ref{fig: 13CO outer radii ratio} shows the ratio \rgas($^{12}$CO)/\rgas($^{13}$CO) for different disk masses. On average, \rgas($^{12}$CO) is 30-45\% larger than \rgas($^{13}$CO), with variations due to \alp, \rc\ or \mdisk being small.

 Note however that we do not consider the effects of isotope-selective photodissociation, which become relevant in the outer part of the disk (cf. \citealt{Miotello2014}).
\section{Measuring \rgas\ from peak intensity maps}
\label{app: moment 0 vs moment 8}

\begin{figure}
    \centering
    \begin{subfigure}{0.99\columnwidth}
    \includegraphics[width=\columnwidth]{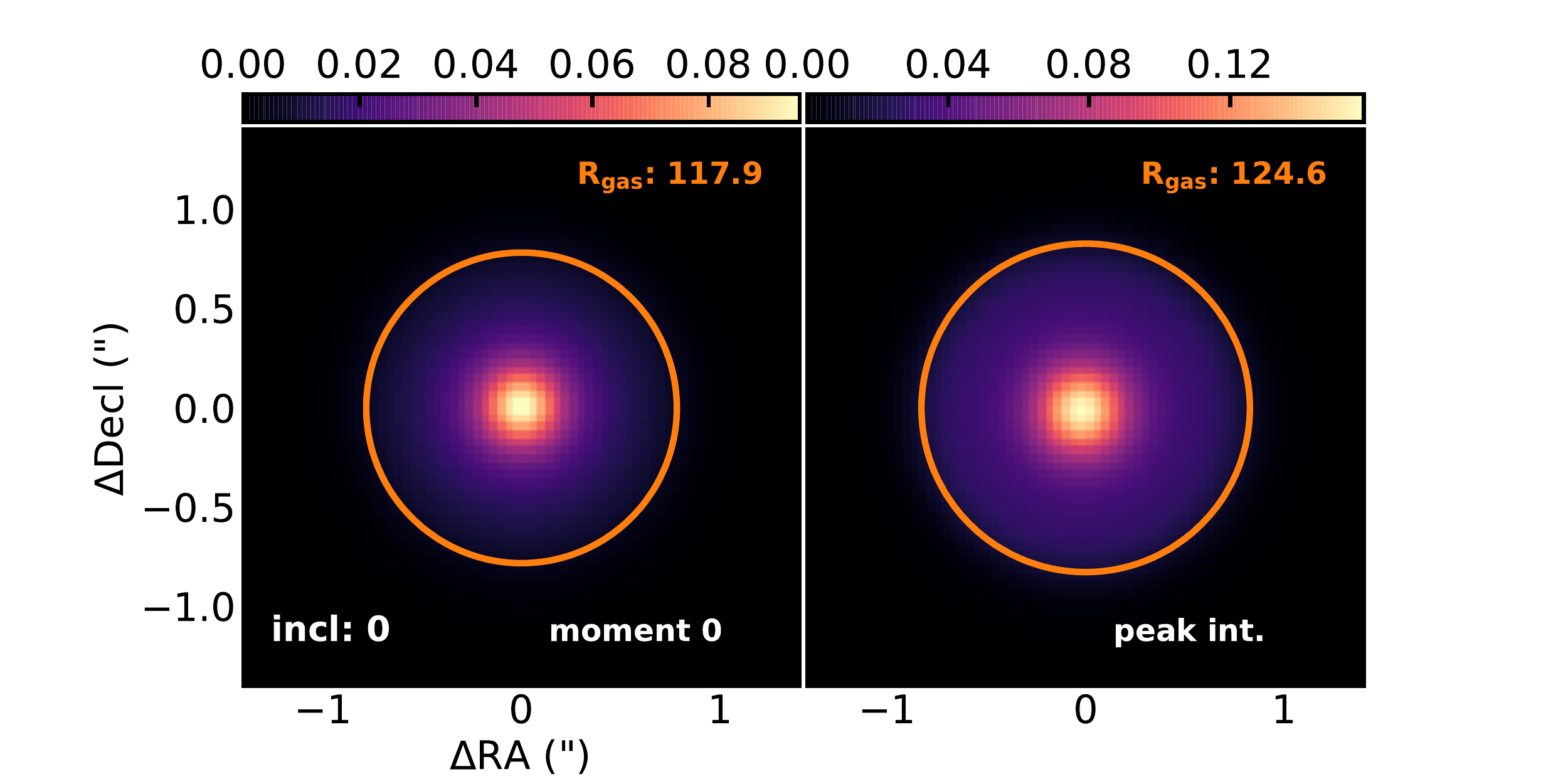}
    \end{subfigure}
    \begin{subfigure}{0.99\columnwidth}
    \includegraphics[width=\columnwidth]{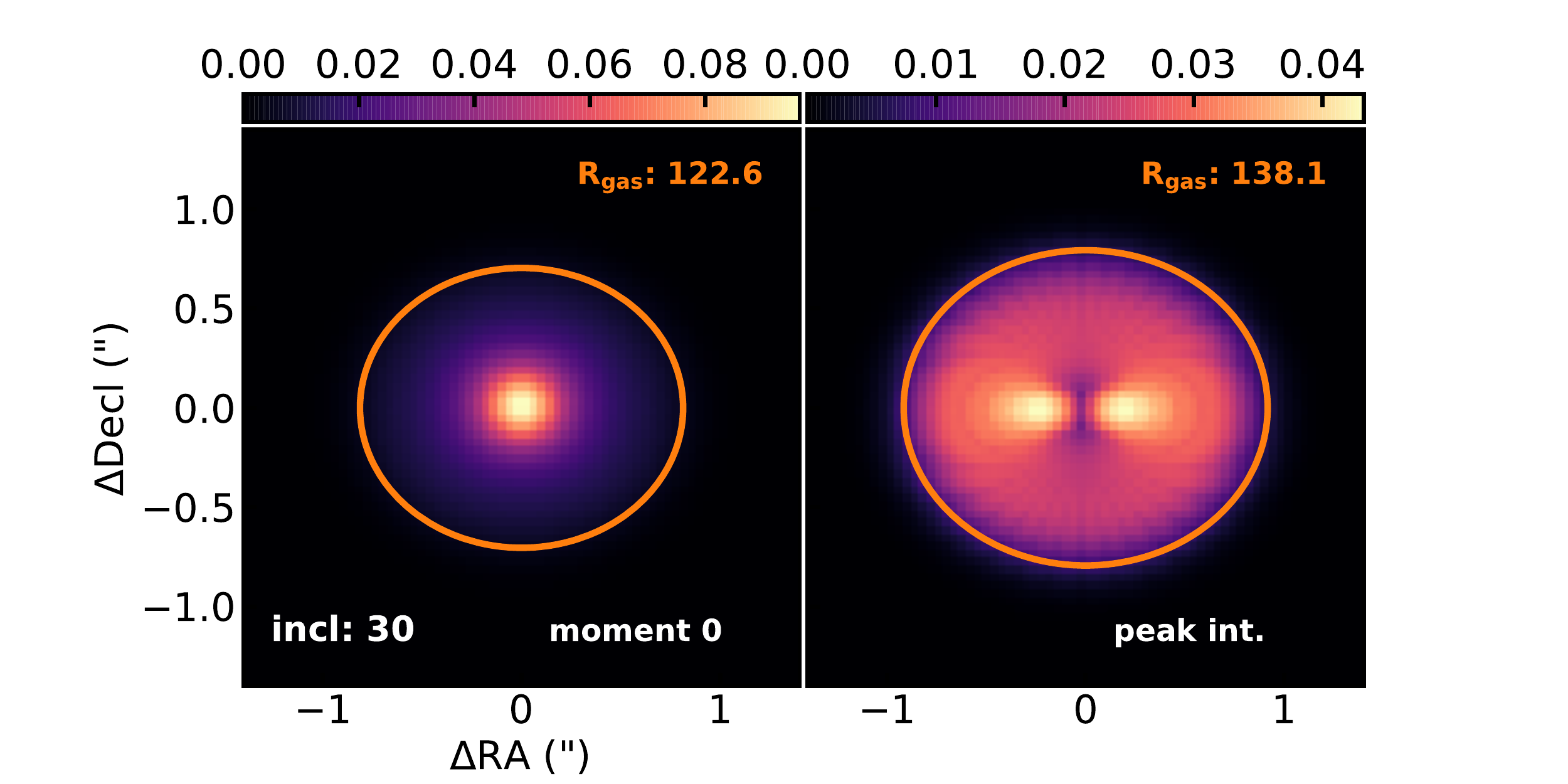}
    \end{subfigure}
    \begin{subfigure}{0.99\columnwidth}
    \includegraphics[width=\columnwidth]{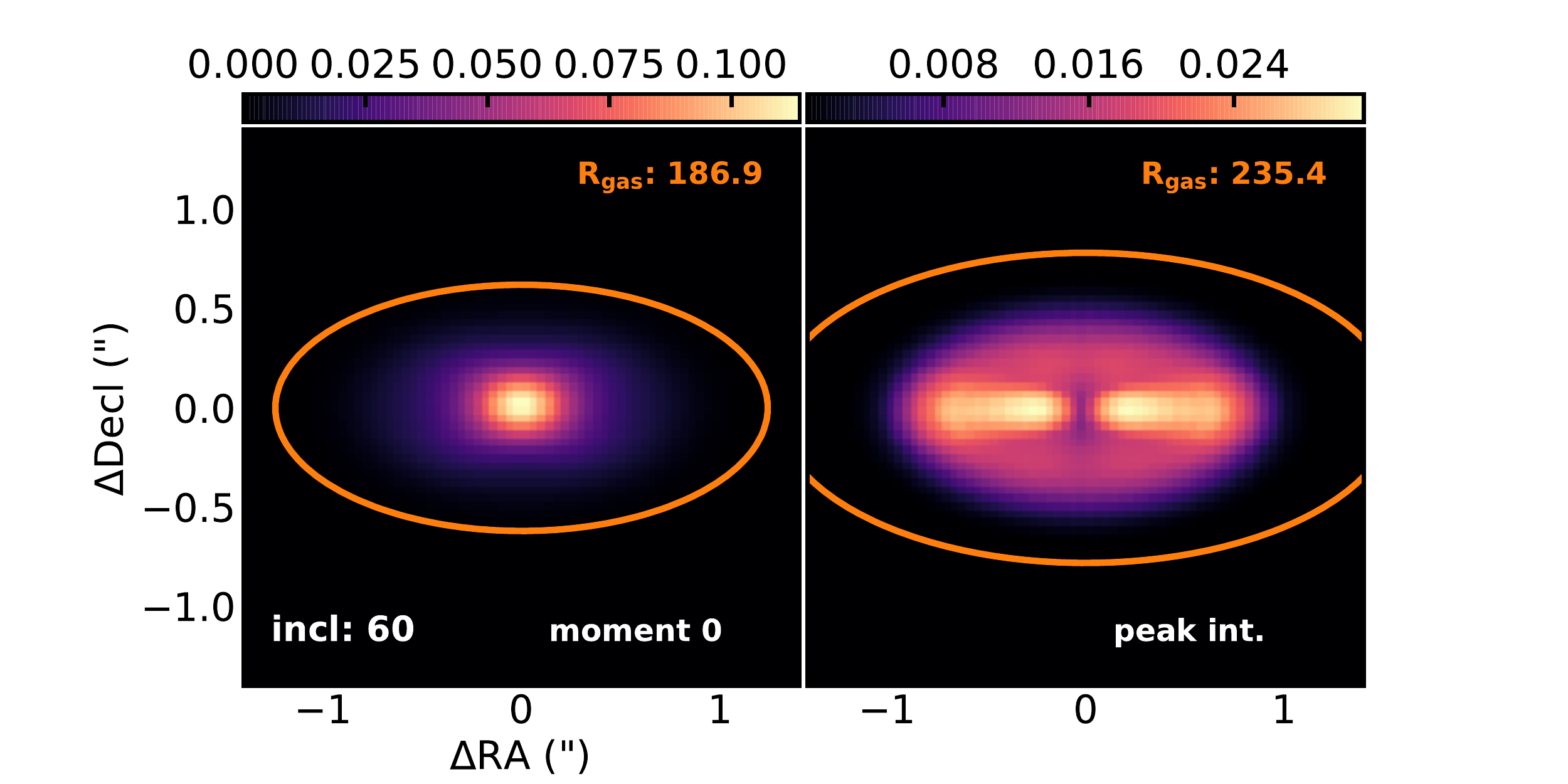}
    \end{subfigure}
    \caption{\label{fig: mom0 vs mom8} A comparison between outer radii derived from $^{12}$CO moment 0 map (left) and the \textbf{peak intensity (moment 8)} map (right). The model shown has $\mdisk = 10^{-3}\ \mathrm{M}_{\odot}$, $\rc=50$ AU and $\alp=10^{-4}$.  }
\end{figure}

The extent of the gas emission is measured from the moment zero map, which is constructed by integrating in spectral image cube over the frequency axis. As a result the gas emission has units [Jy/beam km/s]. This method places extra emphasis on the inner part of the disk, where the Keplerian velocity structure of the gas produces the widest line profiles (in velocity). 

Another method would be to measure the gas radius using the peak intensity map, which is the intensity at peak velocity. This map has units identical to the continuum emission ([Jy/beam]). Compared to the moment 0 map more weight is placed in the outer parts of the disk, moving \rgas\ outward. By removing the dependence on the line width the peak intensity map might also be less affected by inclination (cf. Section \ref{app: inclination}).

\begin{figure}
    \centering
    \begin{subfigure}{0.8\columnwidth}
    \includegraphics[width=\columnwidth]{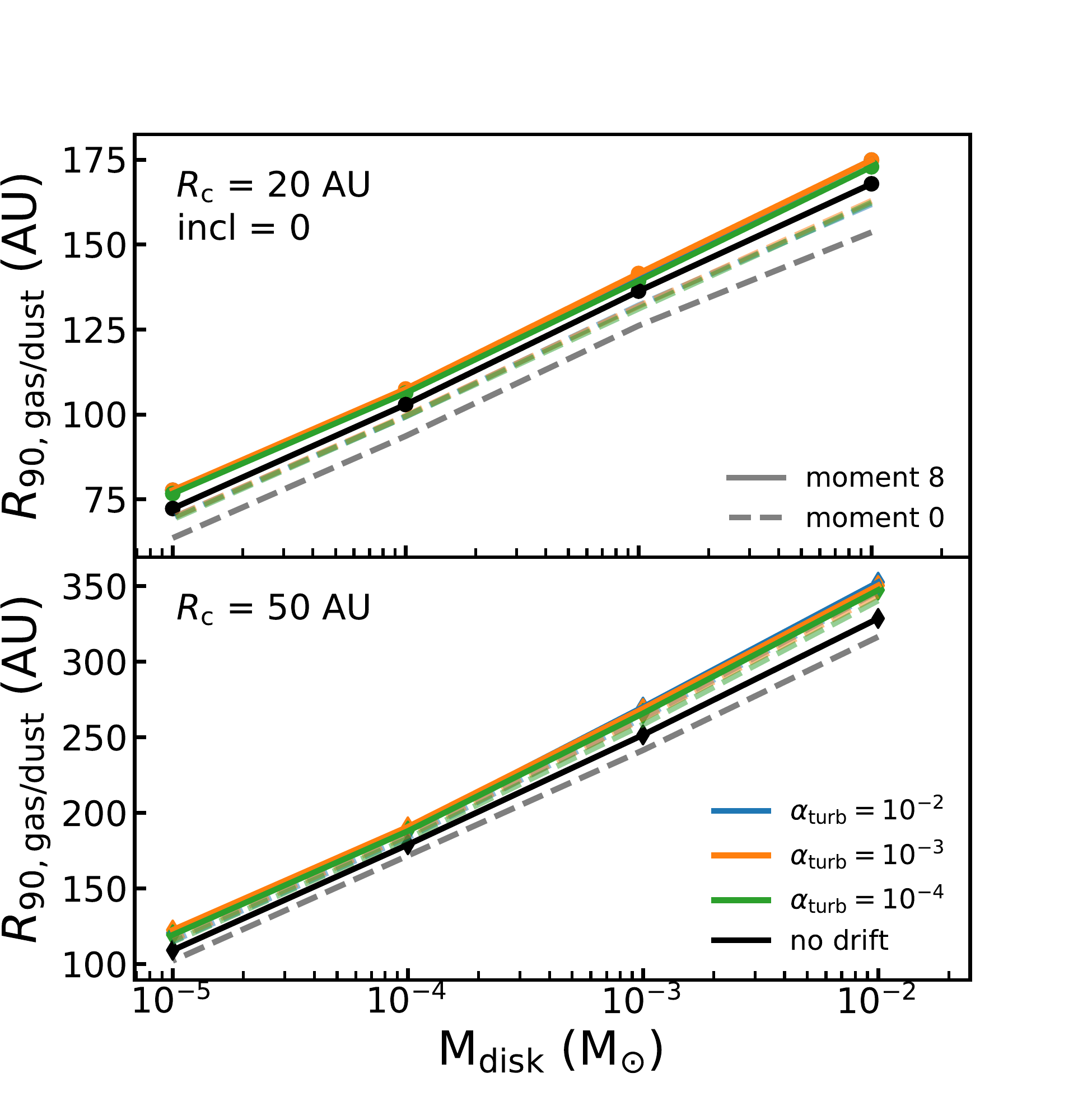}
    \end{subfigure}
    \begin{subfigure}{0.8\columnwidth}
    \includegraphics[width=\columnwidth]{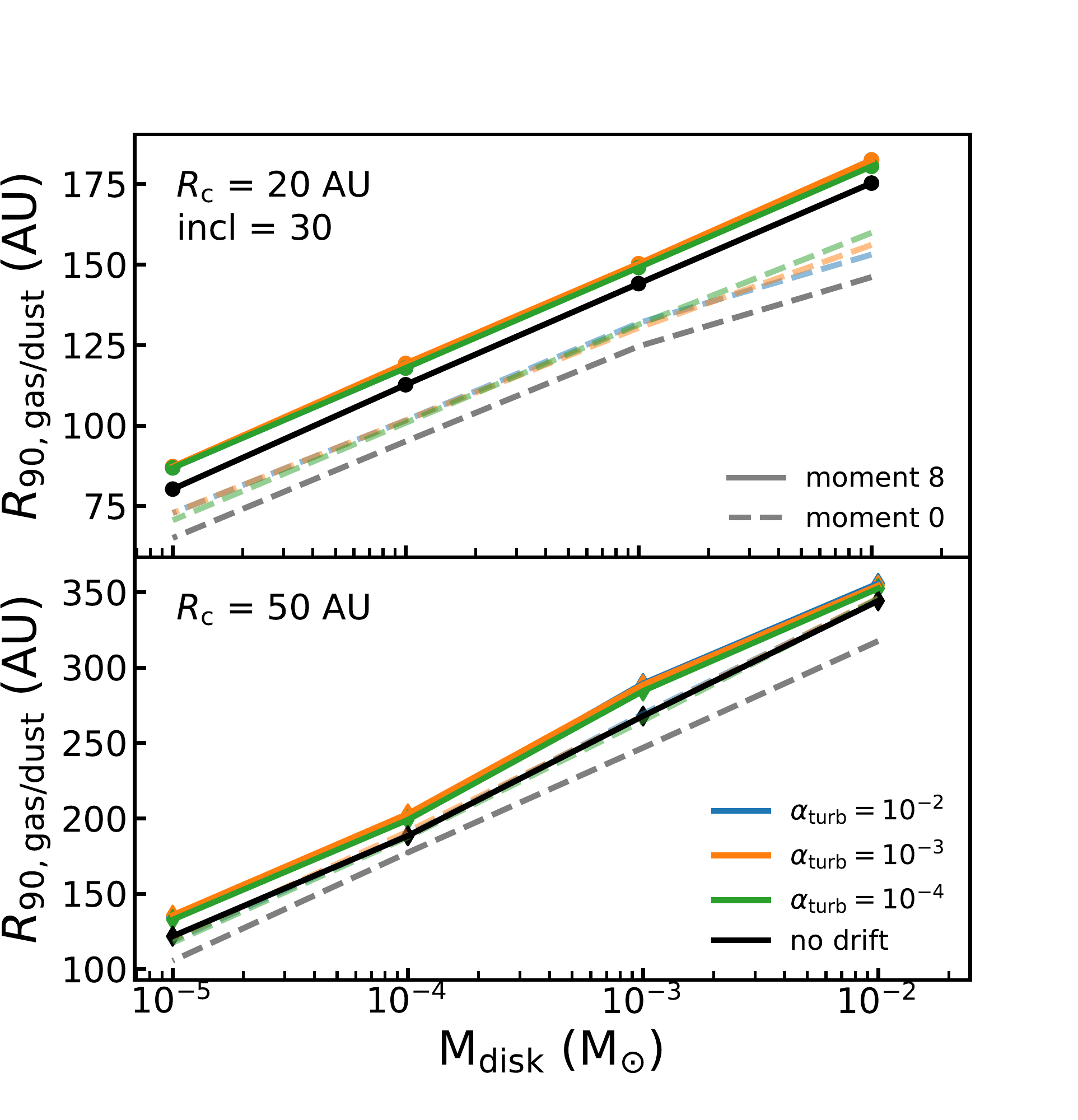}
    \end{subfigure}
    \begin{subfigure}{0.8\columnwidth}
    \includegraphics[width=\columnwidth]{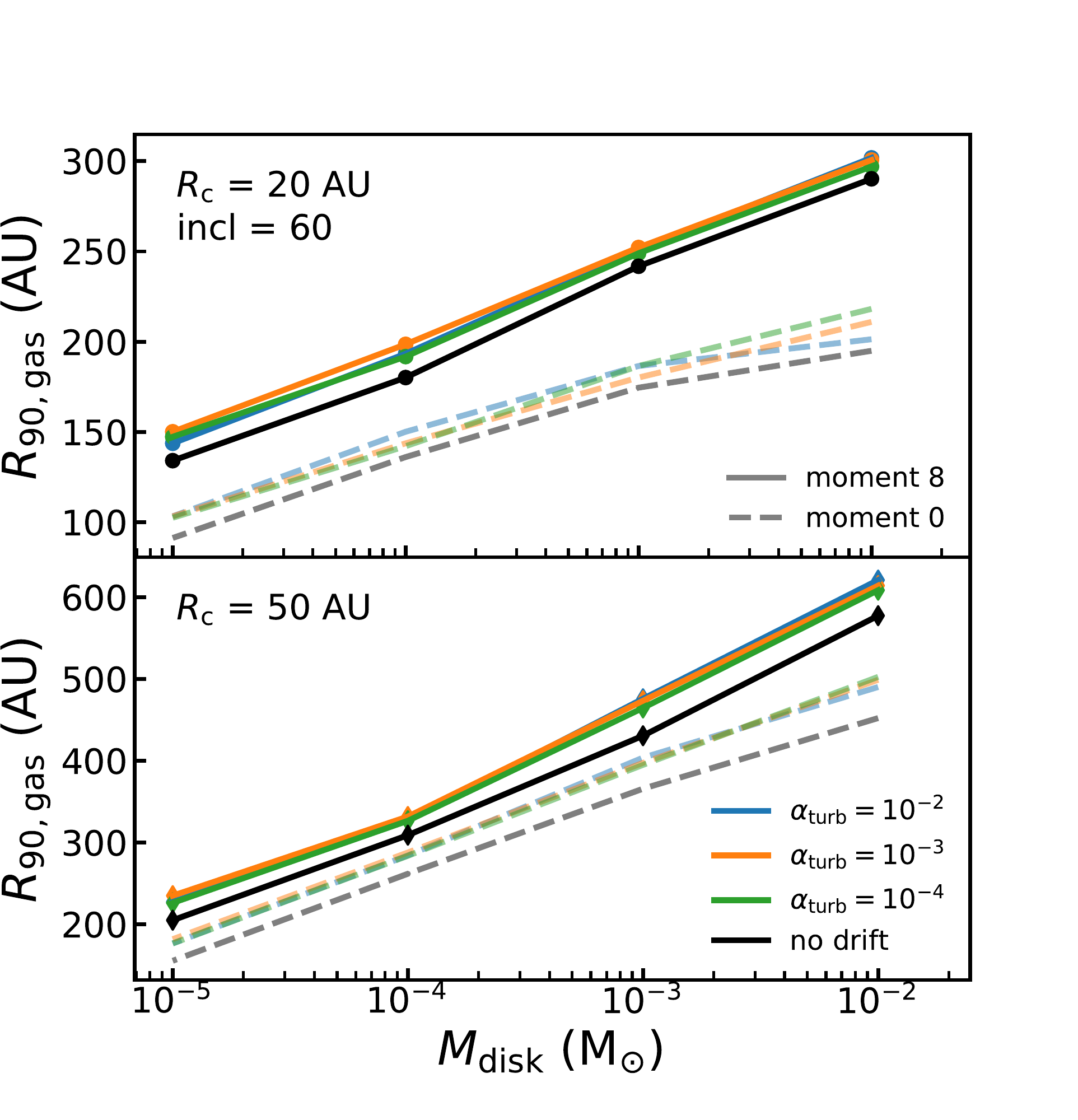}
    \end{subfigure}
    \caption{\label{fig: mom8 vs mass} Gas radii measured from the peak intensity map versus disk mass. The top, middle and bottom figures show disks with inclinations of 0, 30 and 60 degrees, respectively}
\end{figure}

Figure \ref{fig: mom0 vs mom8} shows a comparison between gas outer radii derived from the moment 0 map (shown on the left) and the peak intensity map (shown on the right) for three different inclinations. For the inclined disks ($i=30,60^{\circ}$), there is indeed more emission in the outer disk for the peak intensity map. In all cases \rgas(peak int.) is larger than the \rgas(mom 0). 

In Figure \ref{fig: mom8 vs mass} \rgas(peak int.) and \rgas(mom 0) are compared as function of disk mass and inclination. Over the mass range examined here there we find that \rgas(peak int.) $>$ \rgas(mom 0). However, apart from this offset, \rgas(peak int.) follows the same trend with disk mass as \rgas(mom 0) and is also similarly affected by inclination.
\section{Continuum intensity profiles for R$_c = 20$ AU}
\label{app: dust profiles rc 20}

\begin{figure}
    \centering
    \includegraphics[width=\columnwidth]{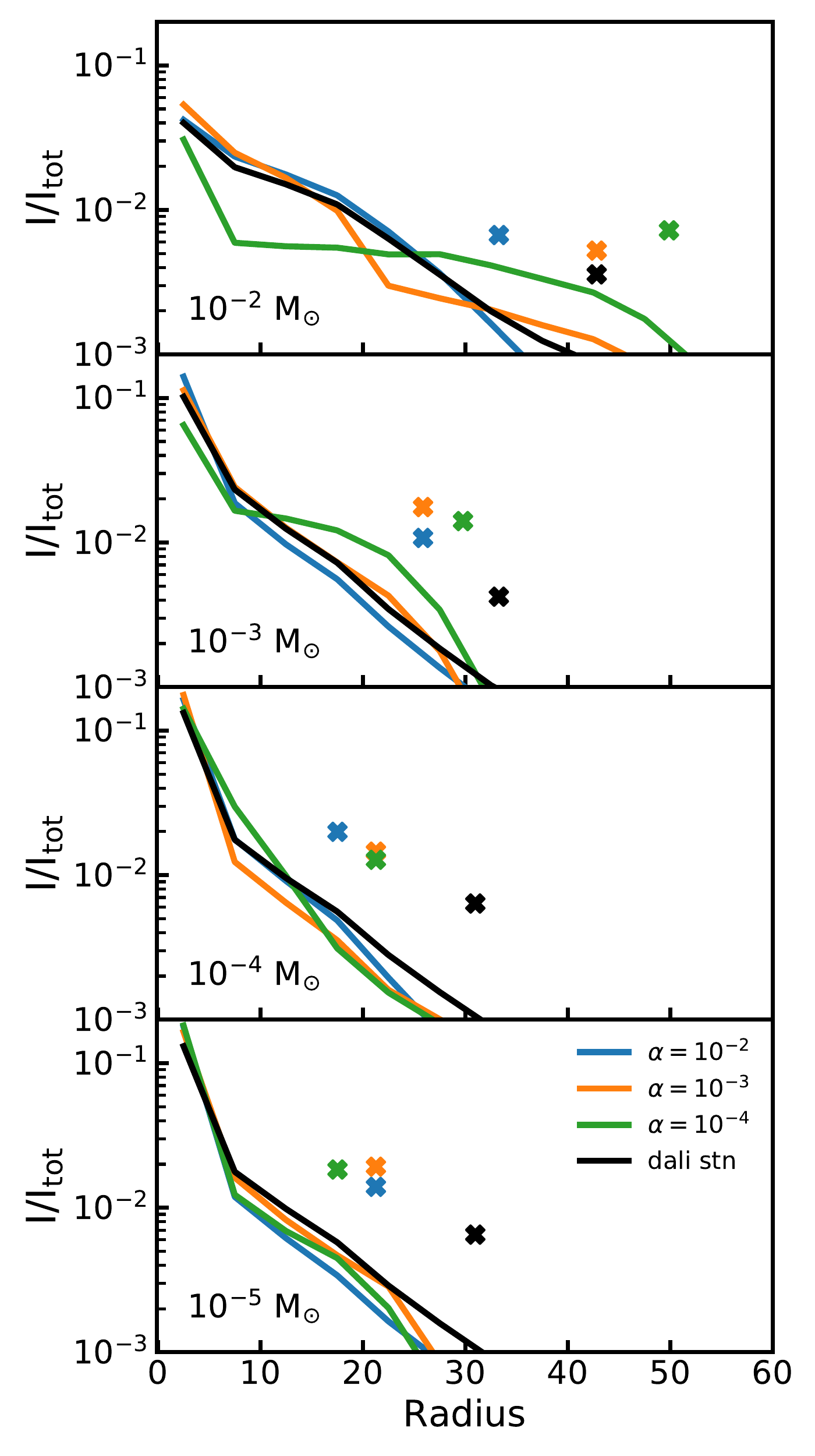}
    \caption{\label{fig: raw dust profiles rc 20} 1300 $\mu$m radial profiles normalised to the total flux. Crosses above the line denote the radii enclosing $90$\% of the flux (heights of the crosses are arbitrary). }
\end{figure}

\begin{figure}
    \centering
    \includegraphics[width=\columnwidth]{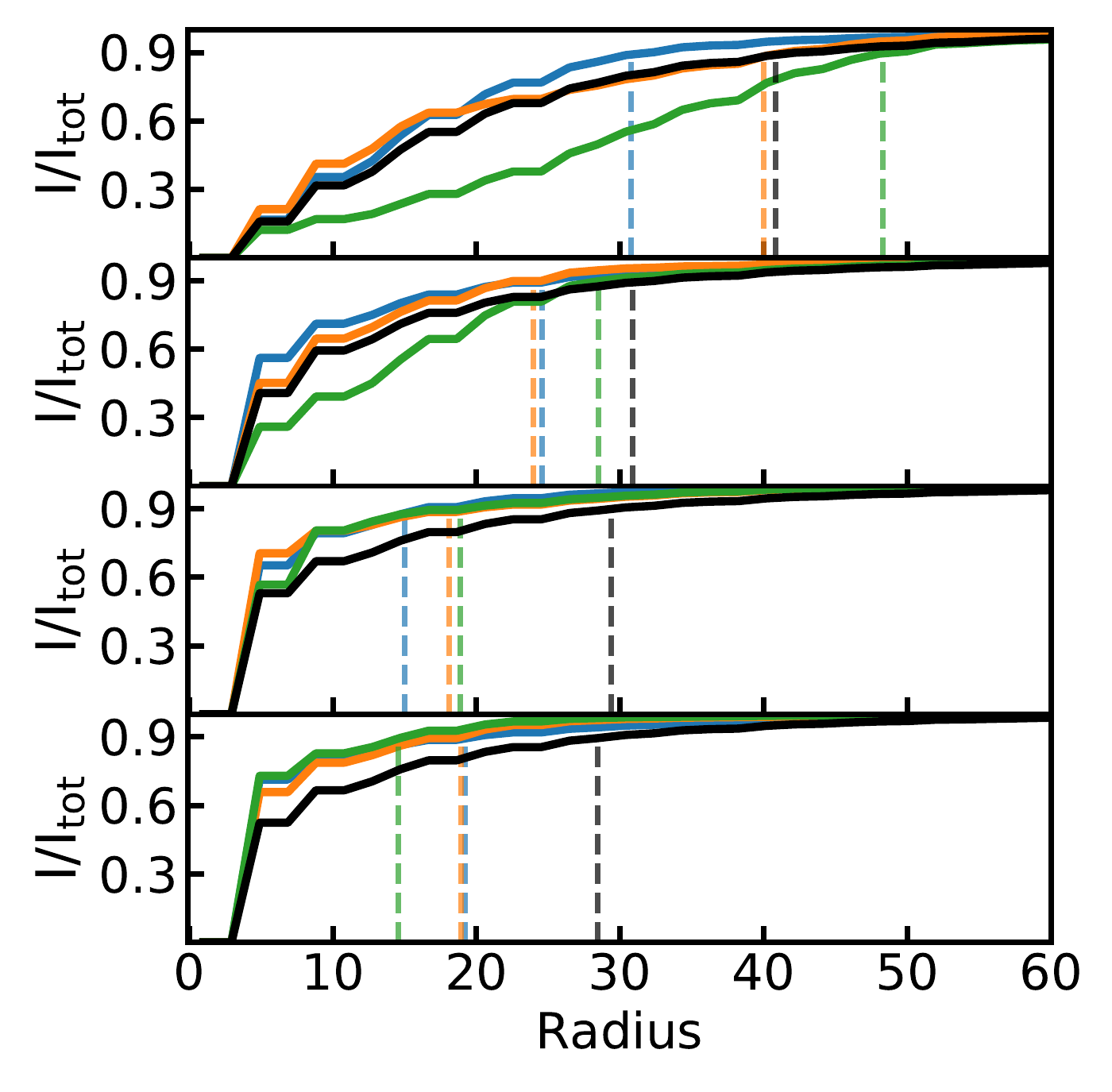}
    \caption{\label{fig: raw dust curves rc 20} 1300 $\mu$m curve-of-growth of the profiles seen in Figure \ref{fig: raw dust profiles rc 20}. Dashed vertical line denote the radii enclosing $90$\% of the flux. }
\end{figure}
\section{Curve-of-growths for the $\rc = 50$ AU dust profiles}
\label{app: dust profiles and curves}

\begin{figure}
    \centering
    \includegraphics[width=\columnwidth]{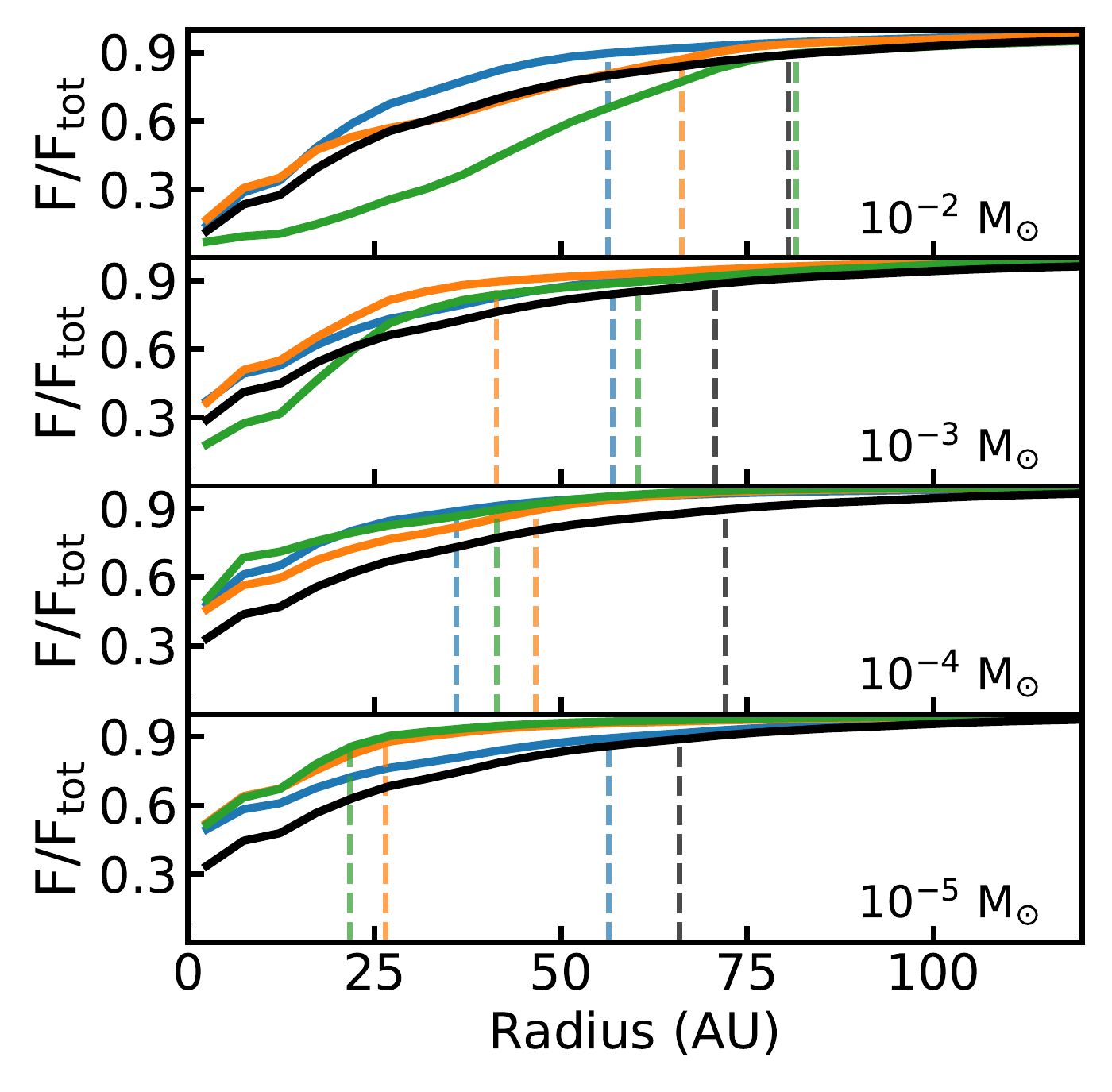}
    \caption{\label{fig: raw dust curves} 1300 $\mu$m curve-of-growth of the profiles seen in Figure \ref{fig: dust profiles}. Dashed vertical line denote the radii enclosing $90$\% of the flux. }
\end{figure}
\section{Deriving a relation between \rgas\ and the CO column density}
\label{app: rgas-rco derivation}

In Section \ref{sec: gas radial profiles} the $^{12}$CO emission profile was found to quickly drop off at a point very close to where the CO column density ($N_{\rm CO}$) drops below $10^{15}\ \mathrm{cm}^{-2}$. Defining \rco\ as the radius where $N_{\rm CO} = 10^{15}\ \mathrm{cm}^{-2}$, we can derive an analytical relation between \rgas\ and \rco.
 
Using equation \eqref{eq: outer radius}, \rgas\ is defined as
\begin{align} 
 0.9 &= \frac{2\pi}{F_{\rm tot}} \int^{\rgas}_{0} I_{\rm CO}(r')r' \mathrm{d}r'\\
     & = \frac{\int^{\rgas}_{0} I_{\rm CO}(r')r' \mathrm{d}r'}{\int^{\rco}_{0} I_{\rm CO}(r')r' \mathrm{d}r'}.
\end{align}
Here we have used the fact that \rco\ effectively encloses all of the $^{12}$CO flux (cf. Figure \ref{fig: gas profiles}). If we assume that the $^{12}$CO emission is optically thick, $I_{\rm CO}(R) = \frac{2\nu^2}{c^2}k_{\rm B} T(R)$, the equation can be rewritten
\begin{equation}
0.9 = \frac{\int^{\rgas}_{0} T(r')r' \mathrm{d}r'}{\int^{\rco}_{0} T(r')r' \mathrm{d}r'},
\end{equation}
where $T_{\rm gas}$ is the gas temperature in the CO emitting layer. The temperature profile can often be well described by a powerlaw, $T(R) = T_c (R/\rc)^{-\beta}$, which can be substituted in the integrals 
\begin{align}
0.9 & = \frac{\int^{\rgas}_{0} \left(\frac{r'}{R_c}\right)^{-\beta}r' \mathrm{d}r'}{\int^{\rco}_{0} \left(\frac{r'}{R_c}\right)^{-\beta}r' \mathrm{d}r'} \\
    & = \frac{\left[\frac{1}{2-\beta} r'^{2-\beta}  \right]^{\rgas}_{0}}{\left[\frac{1}{2-\beta} r'^{2-\beta} \right]^{\rco}_{0}}\\
    &=  \frac{\rgas^{2-\beta}}{\rco^{2-\beta}},
\end{align}
where we have assumed $0 < \beta < 2$. 

We find that \rgas\ and \rco\ are related through
\begin{equation}
\rgas = 0.9^{\frac{1}{2-\beta}}\rco = f^{\frac{1}{2-\beta}}\rco,
\end{equation}
where $f$ represents a more general case where the gas outer radius is defined using a flux fraction $f$.

\section{\ratgasdust vs beamsize and peak SNR for all disk masses}
\label{app: beamsize and peak SNR onepage}

\begin{figure*}
    \centering
    \includegraphics[width=\textwidth]{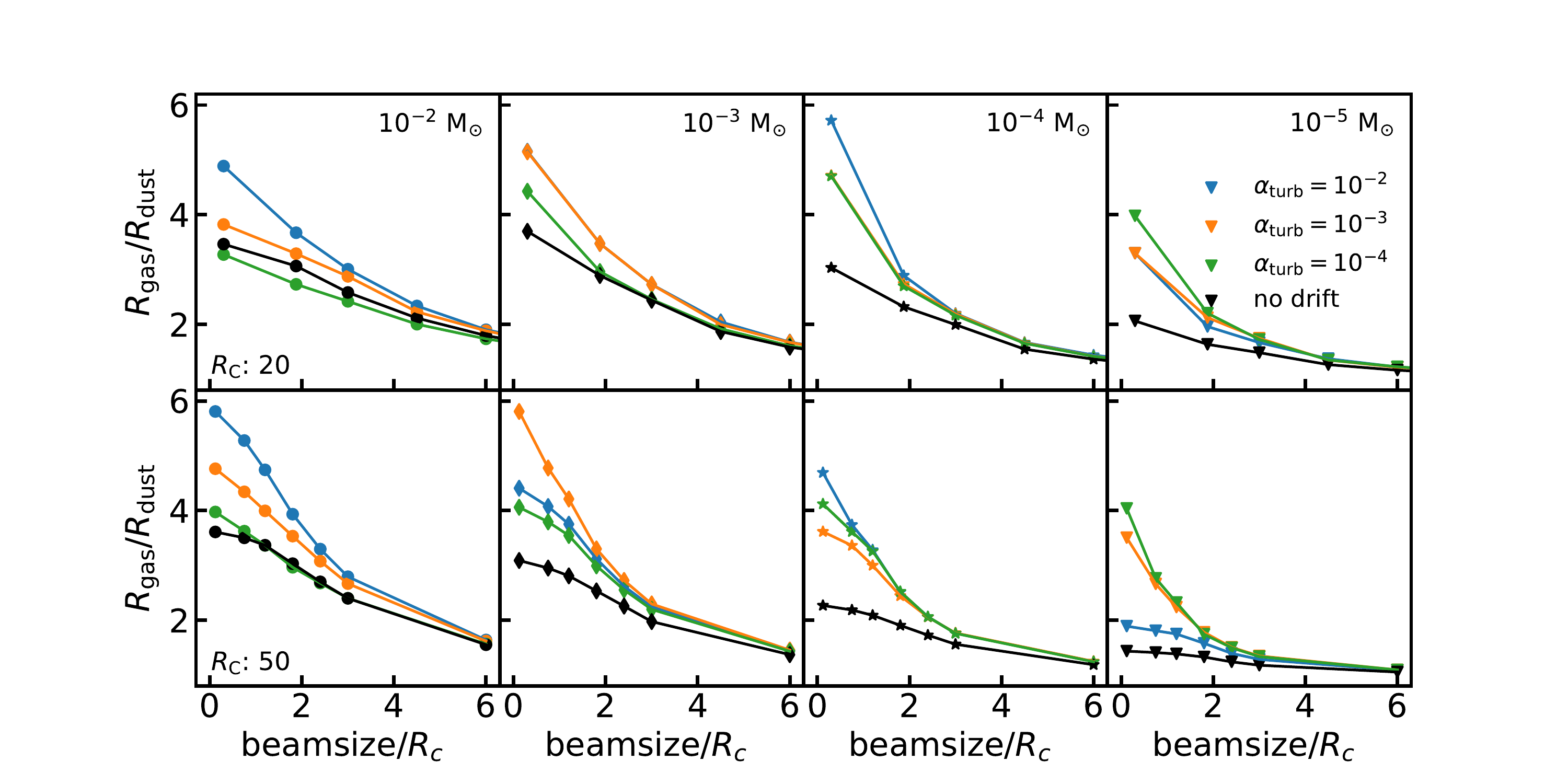}
    \caption{\label{fig: rgas over rdust versus beamsize onepage} \ratgasdust versus beamsize. The effect of the beam scales with its relative size compared to the size of the disk. To highlight this, the beamsize is expressed in terms of the characteristic size of the disk.}
\end{figure*}

\begin{figure*}
    \centering
    \includegraphics[width=\textwidth]{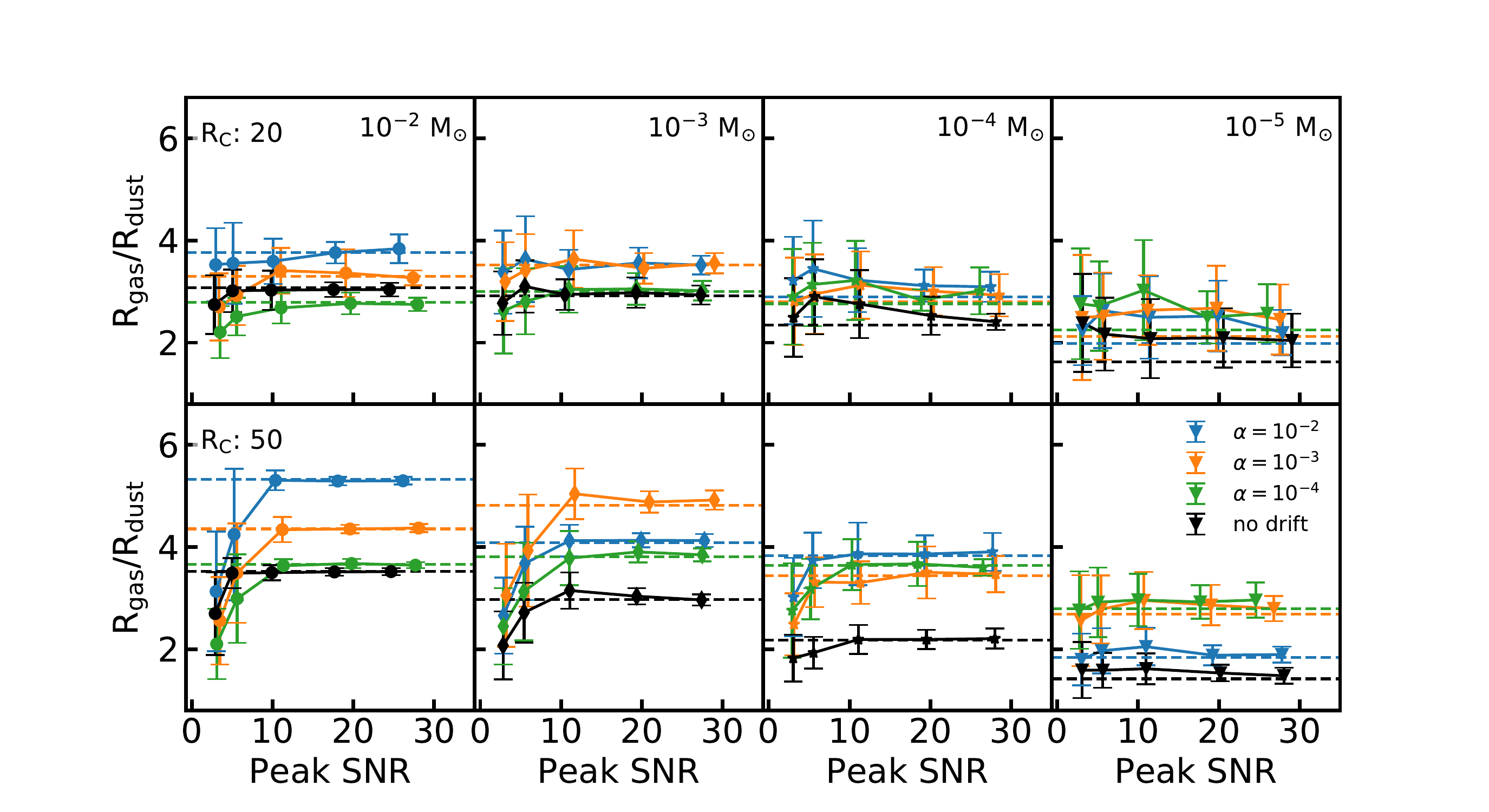}
    \caption{\label{fig: rgas over rdust versus SNR onepage} \ratgasdust versus peak SNR in the moment 0 map of the $^{12}$CO. }
\end{figure*}

\section{Gas radii vs peak SNR}
\label{app: gas radii vs peak snr}

\begin{figure*}
    \centering
    \includegraphics[width=\textwidth]{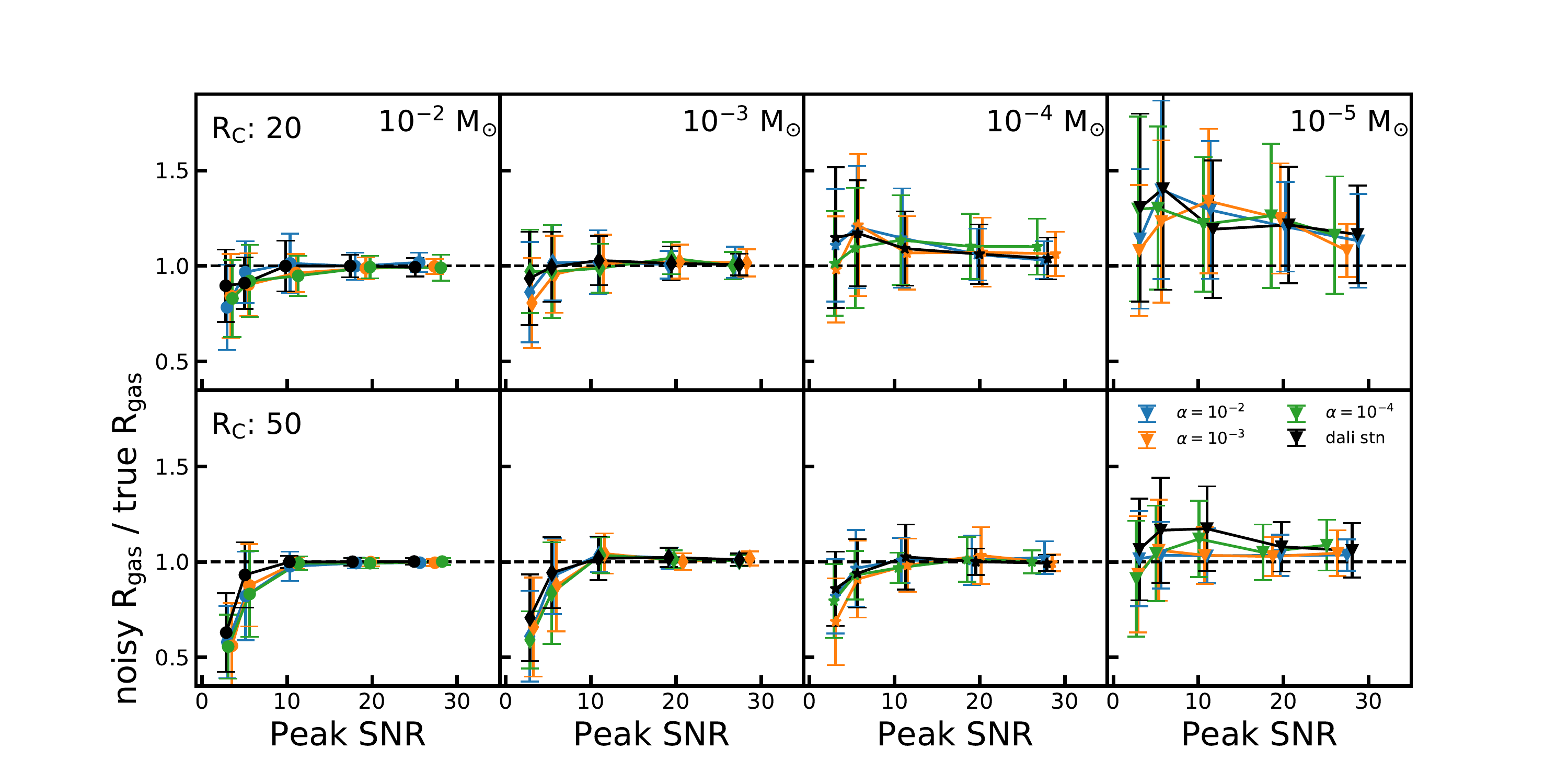}
    \caption{\label{fig: rgas versus SNR} \rgas\ versus peak SNR, normalised to the \rgas\ obtained from the noiseless case. }
\end{figure*}

\section{Mass fractions and flux fractions for the remaining disks}
\label{app: mass fractions}

\begin{figure}
    \centering
    \includegraphics[width=\columnwidth]{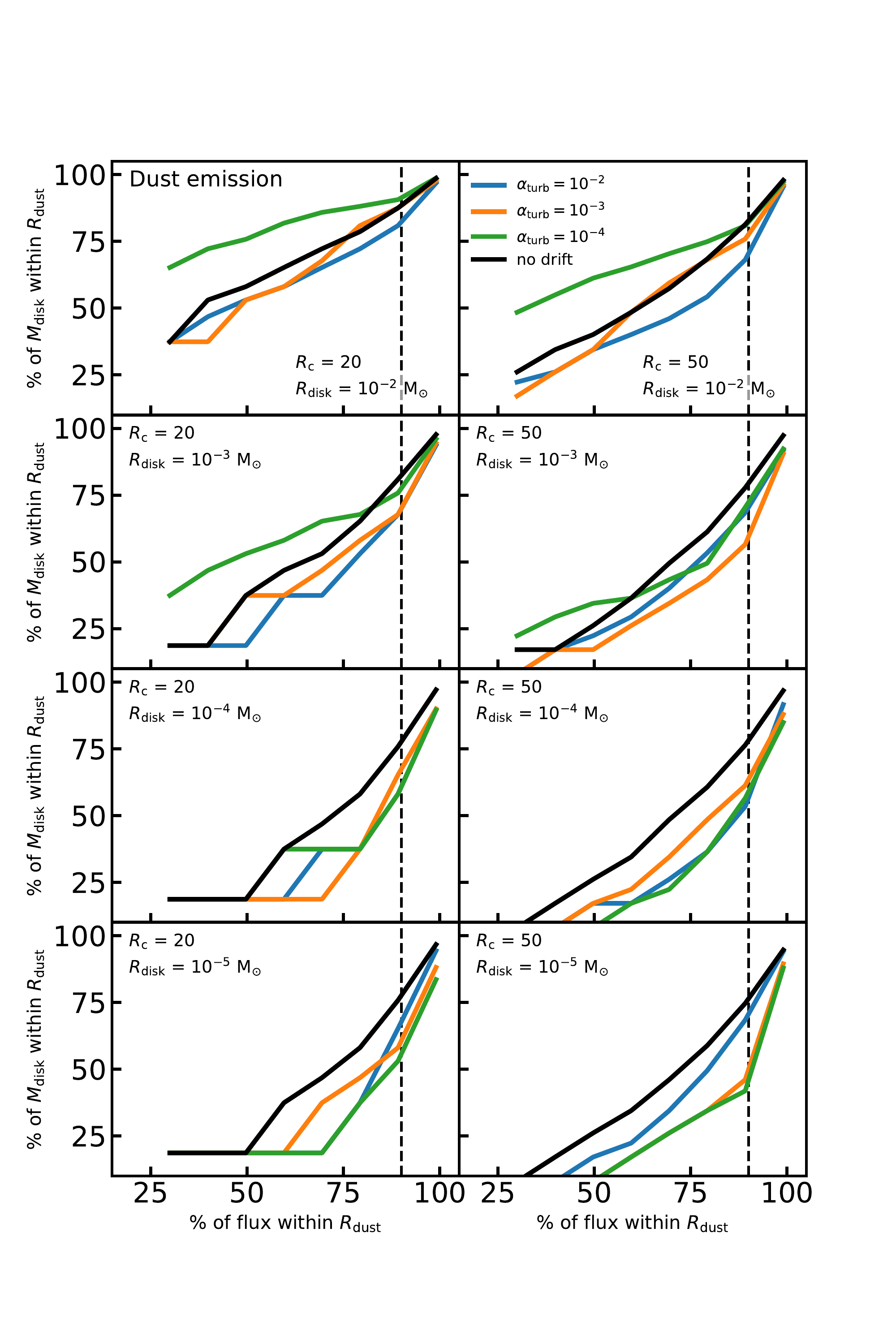}
    \caption{\label{fig: mass fractions dust}  Fraction of continuum flux used to calculate \rdust\ compared to the fraction of \mdisk\ within \rgas. Dashed vertical line indicates 90 \% of the flux.}
\end{figure}

\begin{figure}
    \centering
    \includegraphics[width=\columnwidth]{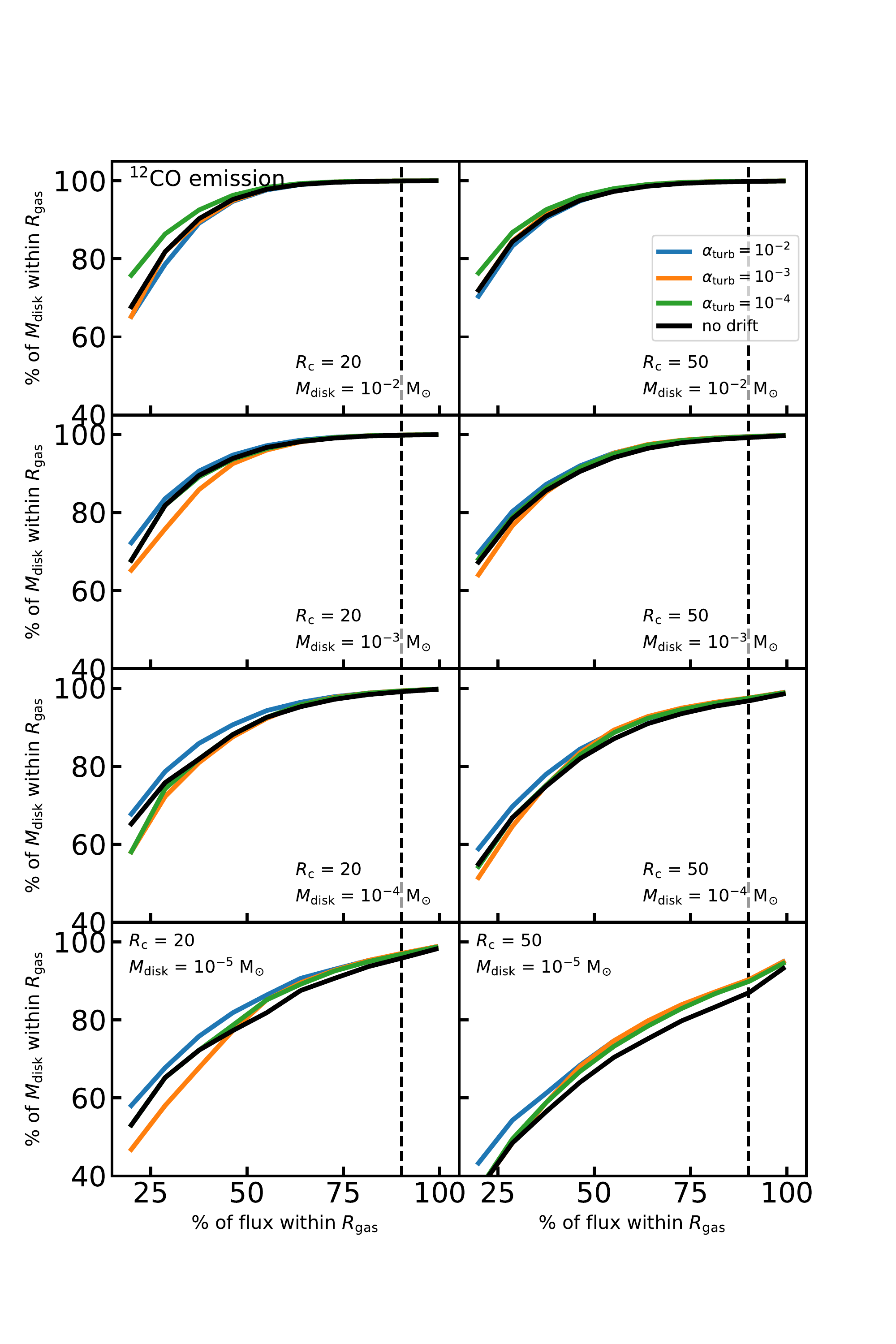}
    \caption{\label{fig: mass fractions gas}  Fraction of $^{12}$CO flux used to calculate \rgas\ compared to the fraction of \mdisk\ within \rgas. Dashed vertical line indicates 90 \% of the flux.}
\end{figure}

\end{appendix}

\end{document}